\documentclass[journal,review,submit,pdftex,moreauthors]{Definitions/mdpi} 
\usepackage{epstopdf}
\firstpage{1} 
\pubvolume{1}
\issuenum{1}
\articlenumber{0}
\pubyear{2026}
\copyrightyear{2026}
\datereceived{ } 
\daterevised{ } % Comment out if no revised date
\dateaccepted{ } 
\datepublished{ } 
\Title{Dense Matter and Compact Stars in Strong Magnetic Fields}

\Author{Monika Sinha $^{1,}$*\orcidA{} and Vivek Baruah Thapa $^{2,}$*\orcidB{}}

\AuthorNames{Firstname Lastname, Firstname Lastname and Firstname Lastname}

\address{%
$^{1}$ \quad Department of Physics, Indian Institute of Technology Jodhpur, Jodhpur 342037, India; ms@iitj.ac.in\\
$^{2}$ \quad Department of Physics, Birangana Sati Sadhani Rajyik Vishwawavidyalaya, Golaghat, Assam 785621, India; vivekthapa@bssrv.ac.in}

\corres{Correspondence: ms@iitj.ac.in (MS); vivekthapa@bssrv.ac.in (VBT)}

\abstract{Compact stars serve as natural systems where matter exists at densities far beyond those achievable in laboratory experiments. Among them, magnetars are expected to possess interior magnetic fields that may reach values of the order of $10^{17}$–$10^{18}$ G. 
These extreme conditions are expected to alter the microscopic and macroscopic properties of dense matter. In this review, we examine how strong magnetic fields affect fermionic matter through mechanisms such as Landau quantization and anomalous magnetic moment interactions. We further discuss the behaviour of magnetized hadronic matter within relativistic mean-field approaches and consider the possible emergence of additional degrees of freedom, including hyperons, $\Delta$ resonances, meson condensates, and quark matter. The consequences of these effects for neutron-star structure and observational constraints are also briefly outlined.}


\begin{document}

%%%%%%%%%%%%%%%%%%%%%%%%%%%%%%%%%%%%%%%%%%
%\setcounter{section}{-1} %% Remove this when starting to work on the template.
%\section{How to Use this Template}

%The template details the sections that can be used in a manuscript. Note that the order and names of article sections may differ from the requirements of the journal (e.g., the positioning of the Materials and Methods section). Please check the instructions on the authors' page of the journal to verify the correct order and names. For any questions, please contact the editorial office of the journal or support@mdpi.com. For LaTeX-related questions please contact latex@mdpi.com.%\endnote{This is an endnote.} % To use endnotes, please un-comment \printendnotes below (before References). Only journal Laws uses \footnote.

% The order of the section titles is different for some journals. Please refer to the "Instructions for Authors” on the journal homepage.

\section{Introduction}\label{intro}
In this review article, we investigate how intense magnetic fields influence the properties of dense matter and the structure of compact stars. 
Neutron stars (NSs) represent extreme astrophysical systems where matter is compressed to supra-nuclear densities while coexisting with intense magnetic fields of order $\sim\,10^{14}-10^{15}$ G \cite{1997ApJ...486L.129V,1998Natur.393..235K,1999ApJ...519L.139W}, naming this special class as \textit{magnetars} \cite{1992ApJ...392L...9D,1992Natur.357..472U,1995MNRAS.275..255T,1996ApJ...473..322T}.
These objects typically have masses close to two solar masses and radii of about ten kilometers, making them ideal environments for investigating the interplay between gravity, dense matter and electromagnetic effects.

Historically, in 1934, Baade and Zwicky \cite{1934PNAS...20..259B} 
predicted the existence of NS composed of neutrons 
immediately after the discovery of neutrons. NSs are 
very compact objects where the degeneracy pressure of relativistic 
neutrons holds the star against its collapse under the influence 
of its own gravity. The observational proof of existence of 
NSs came with the discovery of radio pulsars. The 
first radio pulsar was discovered in 1968 by Hewish \textit{et al.} 
\cite{1968Natur.217..709H}.
Following their initial detection, pulsars were recognized as fast-spinning NSs endowed with intense magnetic fields \cite{1968Natur.218..731G,1969Natur.221...25G}.

\begin{figure}[h!]
    \centering
    \includegraphics[width=6.7cm, keepaspectratio]{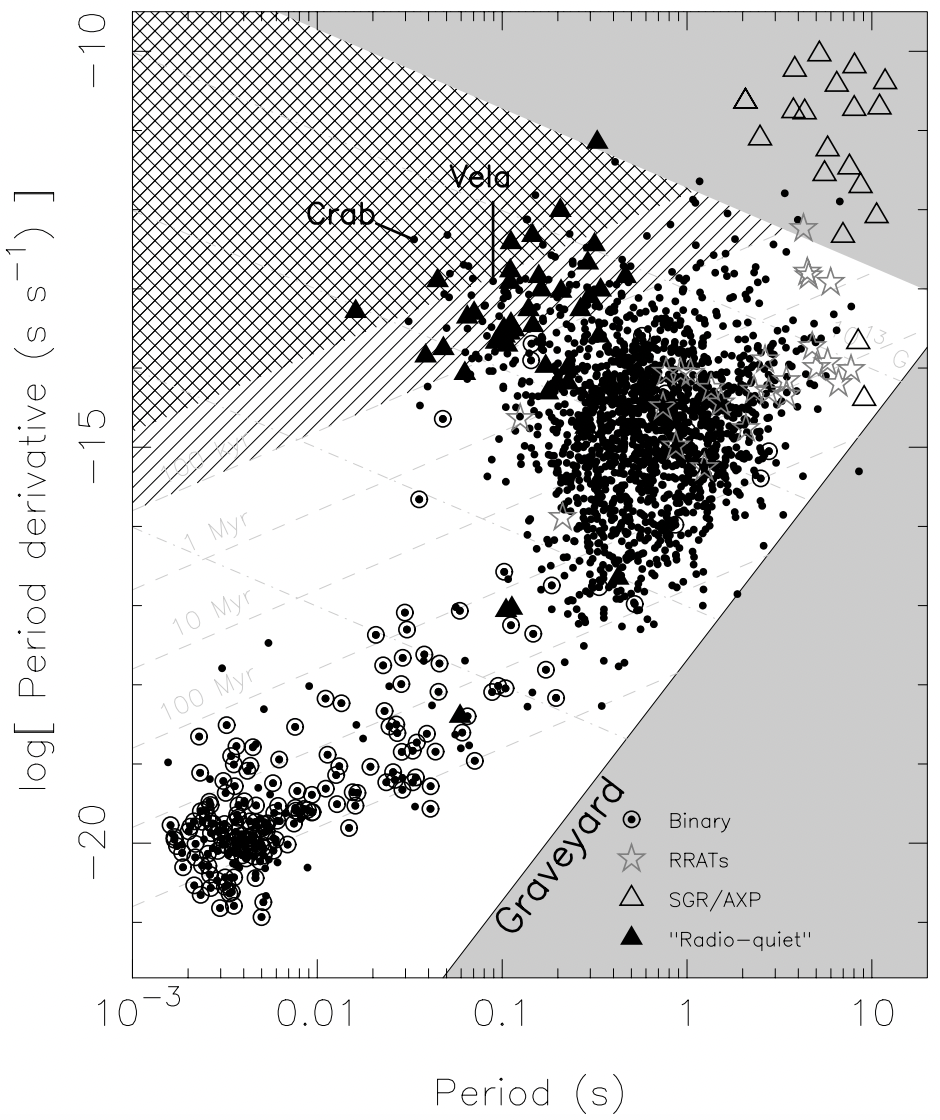}
    \caption{$P–\dot{P}$ diagram of the observed pulsar population. Dashed curves indicate lines of constant characteristic age and surface magnetic field. Binary pulsars, RAdio Transients (RRATs), and magnetars are shown as circles, stars, and upward triangles, respectively. The lower shaded region marks the “pulsar graveyard,” where radio emission is expected to cease, while the upper shaded area corresponds to surface magnetic fields exceeding the quantum critical value of $4 \times 10^{13}$ G. The singly dashed band represents Vela-like pulsars (ages $10-100$ kyr), and the cross-hatched band denotes Crab-like pulsars (ages $<$ 10 kyr). Adapted from Ref. \cite{2004hpa..book.....L}.}
    \label{fig1}
\end{figure}

%Pulsars have very steady but small spin down rate. 
Pulsars are characterized by highly stable rotational periods that gradually increase over time due to energy loss mechanisms. Their observed emission is generally attributed to the conversion of rotational energy, primarily through magnetic braking processes. 
Most radio pulsars 
are single with spin period $0.1-5$ s (see Fig. \ref{fig1}) and the surface magnetic 
field $10^{11}-10^{13}$ G. There is a fraction of radio pulsars 
that rotate with periods of $<10$ ms, and known as millisecond 
pulsars (MSPs). Pulsars in binaries, most of which shows pulse in X-rays, 
also rotate faster than normal pulsars. Both binary pulsars 
and millisecond pulsars are thought to be recycled pulsars which
are spun up accreting matter from their companion stars. 
%For a review on the evolution of pulsars see the 
For a modern review on pulsar evolution and emission properties, see Refs. \cite{2004hpa..book.....L,2006ARA&A..44...17G}
%\cite{1991PhR...203....1B}.
In general, both of them have surface magnetic field $10^8-10^{11}$ G. 
%The radiation from the binary pulsars are triggered by the gravitational energy released by the accreted matter 
In accreting binary systems, the observed radiation is powered by gravitational energy released as matter from the companion star accretes onto the NS
and channeled through the magnetic field lines to produce beamed radiation. 
Hence, in general, for ordinary pulsars the surface magnetic field ranges 
from $10^8-10^{13}$ G.

Soft gamma repeaters (SGRs) \cite{1979Natur.282..587M,1979SvAL....5..343M} and anomalous X-ray pulsars (AXPs) are widely interpreted as manifestations of highly magnetized NSs. 
These sources exhibit recurrent bursting activity with spectral properties \cite{1986Natur.322..152L,1987ApJ...320L.111L} that distinguish them from conventional gamma-ray bursts (GRBs).
Their emission cannot be explained solely through rotational energy loss or accretion processes, and is instead attributed to the release of magnetic energy stored in the stellar interior.
All SGR bursts are softer than GRBs and 
no GRB has been detected so far from the same place of the sky
while SGRs repeat their bursts sporadically. 
%The first AXP was detected in 1981 by Fahlman and Gregory \cite{1981Natur.293..202F}.
AXPs were identified as a distinct class of NS sources in the mid-1990s based on their unique timing and spectral properties \cite{1995ApJ...455..598M,2008A&ARv..15..225M}.
AXPs are characterized by relatively soft X-ray spectra, while their measured spin-down rates are insufficient to account for the observed luminosity.
Consequently, their emission cannot be explained solely by accretion or by rotational energy loss. A significant fraction of SGR and AXP sources are found in association with supernova remnants, supporting their identification as NSs. These objects are generally considered to be relatively young and are characterized by comparatively long spin periods.
The emission from these sources is widely interpreted as originating from the release of magnetic energy stored in the stellar interior \cite{1992ApJ...392L...9D,1996ApJ...473..322T}
%The emission from both SGRs and 
%AXPs were explained as release of the magnetic energy 
with surface magnetic field $\sim 10^{15}-10^{16}$ G, as estimated from spin-down measurements, are classified as {\it magnetars} \cite{2015RPPh...78k6901T,2017ARA&A..55..261K}.

Although the highest value of magnetic field on the surface of neutron 
stars inferred from the observations up to now is $\sim 10^{16}$ 
G, higher field may exist inside the star. 
At present, observational constraints on the internal magnetic field strength and configuration of NSs remain limited. However, theoretical considerations based on energy balance arguments provide useful estimates for the maximum field strength that can be sustained within the star.
For an idealized self-gravitating fluid, the virial theorem can be used to estimate the maximum magnetic field sustainable within the star \cite{1953ApJ...118..116C}.
%For a uniformly distributed, self-gravitating fluid, an estimate of the upper limit on the internal magnetic field can be obtained using the virial theorem. 
In this framework, the equilibrium condition is expressed as
\begin{equation}
2T + W + 3\Pi + {\cal M} = 0,
\end{equation}
where $T$ denotes the rotational kinetic energy, $\Pi$ represents the internal energy, $W$ is the gravitational potential energy, and ${\cal M}$ corresponds to the magnetic energy contribution. 
Now since the kinetic energy and the 
internal energy both are positive, the maximum value of magnetic
energy cannot exceed the gravitational energy of the star.
This argument provides an upper bound on the internal magnetic field strength
as 
\begin{equation}
\frac{4}{3}\pi R^3\,\frac{B_{max}^2}{8\pi} \sim \frac{GM^2}{R},
\end{equation}
which gives 
$B_{max} \sim 2\times 10^8 M/M_\odot (R/R_\odot)^{-2}$ G. For
neutron star (NS), it comes out to be $\sim 10^{18}$ G \cite{1991ApJ...383..745L}. 
However, this is only an estimation. 
%Fully relativistic calculations confirm the simple Newtonian estimates 
More detailed relativistic treatments broadly support these order-of-magnitude estimates obtained from simple considerations \cite{1995A&A...301..757B}.

In recent years, observations from modern space-based instruments have provided increasingly precise constraints on NS properties. In particular, X-ray timing observations by the Neutron Star Interior Composition Explorer (NICER) mission \citep{Miller_2019, Riley_2019, Miller2021} have yielded simultaneous measurements of NS masses and radii through pulse-profile modeling. 
These observations have considerably refined constraints on the dense-matter equation of state (EoS) and provide important benchmarks for theoretical models of magnetized NS matter.

Another major breakthrough in NS physics came with the detection of gravitational waves from binary NS mergers by the LIGO–Virgo collaboration. The first such event, GW170817 \citep{PhysRevLett.119.161101}, provided constraints on the tidal deformability of NSs, thereby placing important limits on the stiffness of the dense-matter EoS. These multi-messenger observations have established a new observational channel for probing NS interiors.

The internal composition of NSs plays a decisive role in determining their observable characteristics. Macroscopic quantities such as mass, radius, and moment of inertia are governed by the underlying EoS, which in turn depends on the properties of dense matter under strong magnetic fields. Furthermore, processes such as thermal evolution and magnetic field decay are closely linked to the microscopic composition of the stellar interior.
Strong magnetic fields influence dense matter through multiple mechanisms. At the microscopic level, they modify the energy spectra of charged particles via Landau quantization and introduce additional contributions through anomalous magnetic moment (AMM) interactions. At larger scales, magnetic stresses affect the equilibrium structure of NSs, leading to modifications in deformation and global properties. A consistent understanding of these effects is therefore essential for interpreting observations of strongly magnetized compact objects.

The remainder of this article is organized as follows. In Sec. \ref{sec:overview} we briefly review the properties and possible composition of dense matter inside NSs. Sec. \ref{sec:magnetized-matter} discusses the basic effects of magnetic fields on fermionic matter, including Landau quantization and AMM interactions. In Sec. \ref{sec:hadronic} we review the behaviour of magnetized hadronic matter within relativistic mean-field (RMF) models and discuss the role of hyperons, $\Delta$-resonances and boson condensates. Sec. \ref{sec:quark} and Sec. \ref{sec:darkmatter} review the implications of magnetic field in quark matter and dark matter, respectively. The implications of magnetic fields for NS structure and observational properties are also addressed in Sec. \ref{sec:NS-structure}. Finally, in Sec. \ref{sec:summary} we summarize the main conclusions and discuss open problems in the study of magnetized dense matter.
We work in natural units, taking $\hbar=c=1$ unless stated otherwise.

\section{Dense matter inside neutron stars}\label{sec:overview}

NSs are formed as remnants of massive stars following core-collapse supernova explosions. These objects typically have masses of about two solar masses and radii of approximately ten kilometers, leading to average densities of the order of $10^{14}$ g cm$^{-3}$. In their interiors, densities may exceed several times the nuclear saturation density, requiring a relativistic description of matter. 
The intense gravitational attraction is counteracted by the degeneracy pressure of fermionic constituents. Structurally, a NS can be broadly divided into distinct layers extending from the surface to the center. The outer crust consists primarily of ions embedded in a sea of electrons, followed by the inner crust where neutron-rich nuclei coexist with free neutrons and electrons. At greater depths lies the core, where matter is compressed to supranuclear densities.
For comprehensive reviews on NS structure and dense matter, see Refs. \cite{2017RvMP...89a5007O,2018RPPh...81e6902B,2023PrPNP.13104041S}.

The physical conditions in these regions differ substantially, and consequently the relevant degrees of freedom may change with increasing density. While the outer layers are relatively well understood within the framework of nuclear physics, the composition of matter in the inner core remains uncertain. At densities several times higher than nuclear saturation density, new particle species or phases of matter may appear depending on the balance between strong interaction dynamics, charge neutrality, and $\beta$-equilibrium. 
Determining which of these possibilities is realized in nature is one of the central problems in the physics of compact stars.

In the core region of a NS, matter is typically dominated by neutrons, with a smaller population of protons and leptons present to ensure charge neutrality and $\beta$-equilibrium.
At sufficiently high densities, the composition of matter in the stellar core becomes increasingly uncertain. A complete theoretical description of matter well above nuclear saturation density is still lacking, largely due to the complexity of strong interactions in this regime and the absence of direct experimental constraints. As a result, a variety of phenomenological models are employed to explore the possible behaviour of dense matter under such conditions.

As the baryon density increases beyond the nuclear saturation density, the Fermi momenta of the constituent particles grow rapidly. Under such conditions it may become energetically favorable for additional degrees of freedom to appear. Several theoretical scenarios have therefore been proposed for the composition of matter in the inner core of NSs. These possibilities arise from different treatments of the strong interaction at supra-nuclear densities and from the requirement of maintaining chemical equilibrium among the constituent particles.
Several theoretical scenarios have been proposed to describe the composition of matter in the inner core of NSs at supra-nuclear densities. These include the following possibilities:

\begin{itemize}

\item {\it Nucleon matter}: 
%With the increase of density the composition of matter does not change. 
In the simplest scenario, the composition remains dominated by nucleons even at higher densities.
It remains nuclear matter with neutrons, protons and electrons (may be with some muons) as its constituent particles. 

\item {\it Heavier strange/ non-strange matter}: With increasing density, the Fermi
momentum of neutrons increases and at certain density it crosses
the threshold energy for creating the heavier baryons or hyperons \cite{PhysRevD.103.063004}.

\item {\it Bosons condensate}: In dense matter the energy of 
pions and kaons may be modified due to interparticle interaction. 
If their energy becomes sufficiently low, Bose condensation
of pion or kaon like excitations can appear 
\cite{1972PhRvL..29..382S,1972PhRvL..29..386S,1986PhLB..175...57K}. 

\item {\it Quark matter}: 
At sufficiently high densities, it is anticipated that baryonic matter may undergo a transition to deconfined quark degrees of freedom
%It is conjectured that at sufficiently high density, the quarks inside the baryons will be deconfined to make deconfined quark matter 
composed of $u$, $d$ and $s$ quarks \cite{1984PhRvD..30..272W,1999paln.conf.....W}.

\item {\textit{Dark Matter:}} In scenarios where dark matter interacts predominantly through gravity, it may accumulate in NSs and coexist with hadronic matter as an additional fluid component \cite{2025MNRAS.543...83K}.
\end{itemize}

In addition there is an idea that whole star is made of strange 
quark matter which is called strange star 
\cite{1999paln.conf.....W,1986ApJ...310..261A,1986NuPhA.450..505F,1986CNPPh..16..289F}. 
Apart from these, there is a possibility that the constituent
particles inside the star may be in a superfluid state. When the
charged particles are superfluid they give rise to superconductivity \cite{2018ASSL..457..401H}.

One of the observables which regulates the EoS is stellar mass. 
%Recent measurements of two pulsars set the lower bound of the maximum mass of NSs 
Recent observations, including precise mass measurements of heavy pulsars and constraints from GW events and NICER observations, have significantly refined the limits on the maximum NS mass.
For PSR J1614-2230, the mass has been accurately measured by Shapiro delay to be 
1.97 $M_\odot$ \cite{2010Natur.467.1081D}, PSR J0348+0432 
has been observed to have mass 2.01 $M_\odot$ \cite{2013Sci...340..448A}
and PSR J0740+6620 with mass $2.08^{+0.07}_{-0.07}~M_\odot$ \cite{2021ApJ...915L..12F}.
Based on atmospheric wind circulation modeling in binary MSP companions, Ref. \cite{2020ApJ...892..101K} reported the mass of PSR J1959+2048 to be $2.18\pm0.09~M_\odot$.
These observations place stringent constraints on acceptable equations of state for dense matter, as the predicted stellar properties depend sensitively on the underlying microscopic description. In particular, the EoS determines the mass–radius relation of compact stars, which can be obtained by solving the Tolman–Oppenheimer–Volkoff (TOV) equations \cite{1939PhRv...55..364T,1939PhRv...55..374O}.

Another recent observation which is worth mentioning here and is 
related to the EoS of NS matter is fast cooling of 
the NS at the center of the youngest supernova remnant 
in Milky way -- Cassiopeia A. With an observation over a decade
it shows very steep decay in its surface temperature \cite{2010ApJ...719L.167H}
which is not compatible with any existing theory of cooling. 
This renders many possibilities of poorly known core compositions 
and the NS inner structure.
These observational developments illustrate how macroscopic properties of NSs are closely connected to the microscopic behaviour of dense matter. The EoS determines not only the mass–radius relation but also influences thermal evolution, neutrino emission processes, and rotational dynamics. Therefore, theoretical models of dense matter must be tested against a variety of astrophysical observations in order to establish a consistent description of NS interiors.

Although the composition of NS matter is largely determined by strong interactions under conditions of $\beta$-equilibrium, the presence of intense magnetic fields can lead to additional modifications. 
In magnetars, magnetic fields may reach strengths at which quantum effects become important for charged particles. Under such circumstances, the magnetic field alters the microscopic properties of matter and can influence both the composition and the thermodynamic behaviour of dense matter. In the following section we therefore discuss the fundamental effects of strong magnetic fields on fermionic matter.

\section{Magnetized Dense matter}\label{sec:magnetized-matter}

The effect of magnetic field on the EoS of dense matter comes 
through its effect on its constituent particles. Whatever be 
the form of highly dense matter inside the NS core 
most of the constituent particles are Fermions. In other possible
scenarios, the most of the constituent particles are Fermions
as stated above. 

\subsection{Fermions in magnetic field}

Under the influence of a magnetic field, the transverse motion of charged fermions is discretized into Landau levels. Relativistic effects become important when the cyclotron energy associated with the magnetic field approaches the rest mass energy of the particle. The corresponding critical field strength, above which such effects cannot be neglected, 
is approximately $B_c^{(e)}=\frac{m^2c^3}{e\hbar}=4.414\times10^{13}$ G for electrons and $B_c^{(p)}=1.487\times10^{20}$ G for protons \cite{1995MNRAS.275..255T}.
Furthermore, magnetic fields interact with the AMMs of baryons, providing an additional mechanism that can substantially modify the EoS at very high field strengths.

\subsubsection{Landau quantization}
When a magnetic field is present, the motion of charged fermions in directions perpendicular to the field is restricted to discrete Landau levels. Adopting a coordinate system in which the magnetic field is aligned along the $z$-axis, the corresponding energy spectrum of the particles is modified. 
The single-particle energy at the $n$-th Landau level can be written as
\begin{equation}
\label{landau}
\epsilon_n = \sqrt{p_z^2 + m^2 + 2 n e |Q| B},
\end{equation}
where $m$ denotes the particle mass, $Q$ is its charge in units of the elementary charge $e$, and $p_z$ represents the momentum component parallel to the magnetic field. 

The allowed Landau levels depend on the spin orientation of the particle. For positively charged fermions, the quantum number $n$ takes values $n=0,1,2,\ldots$ for one spin state, while for the opposite spin orientation it begins from $n=1$. In contrast, for negatively charged particles, the lowest Landau level corresponds to the opposite spin state, with $n=0,1,2,\ldots$, whereas the other spin orientation starts from $n=1$. 
For any species of Fermions the maximum number of occupied Landau levels
are determined by its Fermi momenta as
\begin{equation}
n_{max}=\mathrm {Int}\left(\frac{p_F^2}{2e|Q|{B}}\right).
\end{equation}
As a consequence of Landau quantization, the momentum components in the plane perpendicular to the magnetic field become discretized. With increasing field strength, fewer Landau levels remain populated, indicating a stronger quantization of the system. For a fixed magnetic field, this effect is more significant for particle species with larger Fermi momenta.

The discretization of transverse motion leads to a modification of the phase space structure. At zero temperature, the phase space integral, including spin degeneracy, transforms as
\begin{equation}
2\int_0^{p_F} d^3p \longrightarrow e|Q|B \sum_{n=0}^{n_{max}} (2-\delta_{n,0})
\int_{-p_{F,n}}^{p_{F,n}} dp_z \int_0^{2\pi} d\phi,
\end{equation}
where $\phi$ denotes the azimuthal angle and $p_{F,n} = \sqrt{p_F^2 - 2ne|Q|B}$, with $p_F$ representing the Fermi momentum. In this representation, the continuous integration over transverse momentum is effectively replaced by a discrete sum over the occupied Landau levels.

\subsubsection{Anomalous magnetic moment effect}

In addition to Landau quantization the magnetic field has its
effects on baryons due to their AMM. Unlike Landau quantization
the AMM effect is applicable to all baryons, even if some of
them are charge neutral, as all of them have AMM. The inclusion
of AMM effect modifies the single particle kinetic energy for 
neutral baryons as \cite{2000ApJ...537..351B}
\begin{equation}
\epsilon_s = \sqrt{p_z^2 + \left(\sqrt{m^2+p_\perp^2}+s\kappa B\right)^2},
\end{equation}
and for charged particle as 
\begin{equation}
\epsilon_{n,s} = \sqrt{p_z^2 + (\sqrt{m^2+2ne|Q|B}+s\kappa B)^2},
\end{equation}
where $s\kappa$ is the AMM, $s$ being the spin projection and 
$\kappa$ being the strength of the AMM. 

With these two modifications (from Landau quantization and AMM
effect) in single particle energy and in phase space the presence
of magnetic field affects the energy density, chemical potential
and hence the thermodynamic pressure of dense matter.

\subsection{Energy momentum tensor in magnetic field}

In the presence of an electromagnetic field, the total energy–momentum tensor of a fermionic system can be separated into two distinct contributions: a matter component and a field component. Accordingly, it can be expressed as
\begin{equation}
T^{\mu \nu} = T^{\mu \nu}_m + T^{\mu \nu}_f,
\end{equation}
where the matter part is written as
\begin{equation}
T^{\mu\nu}_m= \varepsilon_m u^\mu u^\nu - P_m(g^{\mu\nu}-u^\mu u^\nu)
+ \frac12 (M^{\mu \lambda}F_\lambda^\nu + M^{\nu \lambda}F_\lambda^\mu),
\label{tmatter}
\end{equation}
Here, $\varepsilon_m$ denotes the energy density of matter, $P_m$ represents the thermodynamic pressure, and $M^{\mu \nu}$ corresponds to the magnetization tensor. This term effectively combines the contribution from an isotropic fluid with an additional correction arising from magnetization effects.

The field contribution to the energy–momentum tensor is given by
\begin{equation}
T^{\mu\nu}_f = -\frac{1}{4\pi}F^{\mu\lambda}F^\nu_\lambda + \frac1{16\pi} g^{\mu \nu} 
F^{\rho\sigma} F_{\rho\sigma}. 
\label{tmag}
\end{equation}

Since our focus is on the influence of magnetic fields in dense matter, we neglect macroscopic electric fields in the bulk medium. Under this assumption, Eqs.~(\ref{tmatter}) and (\ref{tmag}) simplify to \cite{2002PhRvD..65e6001K,2010PhRvD..81d5015H,2019PhRvC..99f5803F}
\begin{eqnarray}
T^{\mu\nu}_m &=& \varepsilon_m u^\mu u^\nu - P_m(g^{\mu\nu}-u^\mu u^\nu)
+ M {B} \left(g^{\mu\nu}-u^\mu u^\nu + \frac{{B}^\mu {B}^\nu}{{B}^2}\right),\\
T^{\mu \nu}_f &=& \frac{{B}^2}{4\pi} \left(u^\mu u^\nu - \frac12 g^{\mu\nu}\right) 
- \frac{{B}^\mu {B}^\nu}{4\pi},
\end{eqnarray}
where ${B}^\mu{B}_\mu = - {B}^2$, and the magnetization per unit volume is defined as $M = -\partial P/\partial B$ with $P$ being the parallel pressure.

In the rest frame of the hadronic fluid, assuming the magnetic field is aligned along the $z$-direction, the corresponding components of the matter and field tensors take the forms
\begin{equation}
\label{eq:Tmatter}
T^{\mu\nu}_m = \left(\begin{array}{cccc} 
                    \varepsilon_m & 0 & 0 & 0\\
                    0 & P_m-MB & 0 & 0\\
                    0 & 0 & P_m-MB & 0\\
                    0 & 0 & 0 & P_m 
               \end{array}\right),
\end{equation}
\begin{equation}
\label{eq:Tfield}
T^{\mu\nu}_f = \frac{B^2}{8\pi} \left(\begin{array}{cccc} 
                    1 & 0 & 0 & 0\\
                    0 & 1 & 0 & 0\\
                    0 & 0 & 1 & 0\\
                    0 & 0 & 0 & -1 
               \end{array}\right).
\end{equation}

In several earlier studies \cite{2013NuPhA.898...43S,2010PhRvC..82f5802F,2011PhRvD..83d3009P}, the total energy density has been written as a direct sum of the matter and magnetic field contributions,
\begin{equation}
\varepsilon = \varepsilon_m + \frac{B^2}{8\pi}.
\label{ener0}
\end{equation}

From Eqs.~(\ref{eq:Tmatter}) and (\ref{eq:Tfield}), it follows that the presence of a magnetic field introduces direction-dependent stresses. The effective pressure perpendicular to the magnetic field is given by
\begin{equation}
P_\perp = P_m - MB + \frac{B^2}{8\pi},
\label{perp}
\end{equation}
while along the direction of the magnetic field the corresponding expression is
\begin{equation}
P_\parallel = P_m - \frac{B^2}{8\pi}.
\label{par}
\end{equation}

Clearly, the presence of a magnetic field introduces anisotropy in the spatial components of the energy–momentum tensor.
%the presence of magnetic field makes the pressure anisotropic. 
The more field is strong the more pressure is
anisotropic. 
%However, later it was shown that the magnetic field does not make the pressure anisotropic, 
However, subsequent studies clarified that the thermodynamic pressure itself remains isotropic, 
it just induces the anisotropy in energy momentum tensor. By definition, pressure is a scalar quantity. Even under the conditions of an intense magnetic field, the pressure remains isotropic because the magnetic field modifies the energy levels of particles (for example, through Landau quantization) but does not change the fundamental thermodynamic definition of pressure. The magnetic field affects the numerical value of pressure through its appearance in the grand potential, but it does not split pressure into parallel or perpendicular components at the thermodynamic level. 
Therefore, pressure remains a scalar thermodynamic quantity.
%Hence, pressure remains a scalar function.
The anisotropicity in the energy momentum tensor is also induced by magnetisation. However its effect is canceled by the Lorentz force associated with the magnetization. For details we refer the readers to the article \cite{2015MNRAS.447.3785C}.

\section{Hadronic matter}\label{sec:hadronic}

Studies on the behaviour of a strongly magnetized electron gas in NS matter have been carried out in several works \cite{1977FCPh....2..203C,1989ApJ...342..958F,1991ApJ...374..652A,1992AnPhy.216...29F,1993ApJ...416..276R}, particularly in connection with the outer regions of NSs where leptonic components play a dominant role. In contrast, the composition of the stellar core is primarily governed by baryonic degrees of freedom in most theoretical descriptions of dense matter. Beyond nucleons, additional constituents such as hyperons, heavier non-strange baryons including $\Delta$ resonances, as well as meson condensates, may emerge at sufficiently high densities.

\subsection{Theoretical model - a generalized relativistic mean field description }
Dense hadronic matter is often described within the framework of RMF theory, originally formulated in the Walecka model \cite{1974AnPhy..83..491W}. This approach has been widely employed due to its ability to reproduce key features of nuclear systems, including saturation properties and scattering observables. In the present work, we adopt a generalized RMF description that incorporates a broad set of constituents relevant for NS matter, such as nucleons, hyperons, $\Delta$ resonances, and meson condensates. 

Within this framework, the interactions among baryons are mediated by meson exchange, typically involving the isoscalar-scalar $\sigma$, isoscalar-vector $\omega$, and isovector-vector $\rho$ fields. A known limitation of the original Walecka model is its tendency to predict an unrealistically high incompressibility for nuclear matter. Several modifications have been proposed to address this issue. One common approach introduces nonlinear self-interaction terms of cubic and quartic order for the scalar field in the Lagrangian \cite{1977NuPhA.292..413B}. An alternative formulation was developed by 
Zimanyi and Moszkowski (ZM) \cite{1990PhRvC..42.1416Z}, where the nonlinearity arises through a modified relation between the effective baryon mass and the scalar field. Another widely used extension involves incorporating density-dependent coupling constants, which allow for a more flexible and realistic description of nuclear interactions \cite{1995PhLB..345..355L,1995PhRvC..52.3043F}.

A large fraction of existing studies on magnetized hadronic matter have been carried out within the framework of nonlinear RMF models. In the presence of a static magnetic field (hereafter denoted as $B$), the total Lagrangian density of the system can be decomposed into matter and field contributions, and is written as
\begin{equation}
\label{eq:Ldens}
{\cal L} = {\cal L}_m + {\cal L}_f,
\end{equation}
where ${\cal L}_m$ represents the matter sector, while ${\cal L}_f$ accounts for the electromagnetic field contribution.

Within density-dependent RMF formulations, the matter part of the Lagrangian density takes the form
\begin{equation}
\begin{aligned}
\mathcal{L}_m & = \sum_{b} \bar{\psi}_b(i\gamma_{\mu} D^{\mu}_{(b)} - m^{*}_b) \psi_b + \sum_{d} \bar{\psi}_{d\nu}(i\gamma_{\mu} D^{\mu}_{(d)} - m^{*}_d) \psi^{\nu}_{d} \\ 
& + \sum_{l} \bar{\psi}_l [i\gamma_{\mu} (\partial^{\mu}+ieQA^\mu) - m_l]\psi_l + D^{(\bar{K})*}_\mu \bar{K} D^\mu_{(\bar{K})} K - m^{*^2}_K \bar{K} K \\
& + \frac{1}{2}(\partial_{\mu}\sigma\partial^{\mu}\sigma - m_{\sigma}^2 \sigma^2) + \frac{1}{2}(\partial_{\mu}\sigma^*\partial^{\mu}\sigma^* - m_{\sigma^*}^2 \sigma^{*2})  \\
& -  \frac{1}{4}\omega_{\mu\nu}\omega^{\mu\nu} + \frac{1}{2}m_{\omega}^2\omega_{\mu}\omega^{\mu} - \frac{1}{4}\boldsymbol{\rho}_{\mu\nu} \cdot \boldsymbol{\rho}^{\mu\nu} + \frac{1}{2}m_{\rho}^2\boldsymbol{\rho}_{\mu} \cdot \boldsymbol{\rho}^{\mu} \\
& - \frac{1}{4}\phi_{\mu\nu}\phi^{\mu\nu} + \frac{1}{2}m_{\phi}^2\phi_{\mu}\phi^{\mu},
\end{aligned}
\label{lagm}
\end{equation}
with the interaction between baryons and mesonic fields incorporated through the covariant derivative,
\begin{equation}\label{eqn.2}
D_{\mu (j)} = \partial_\mu + ig_{\omega j} \omega_\mu + ig_{\rho j} \boldsymbol{\tau}_j \cdot \boldsymbol{\rho}_{\mu} + ig_{\phi j} \phi_\mu + ieQA^{\mu},
\end{equation}
where the index $j$ runs over baryons ($b, d$) as well as the (anti)kaon sector, $A^{\mu}$ is the electromagnetic vector potential, $eQ$ is the
charge of the particle ($e$ being unit `+' charge). The coupling strengths $g_{pj}$ depend explicitly on the density, with the index $p$ labeling the exchanged mesons. The operator $\boldsymbol{\tau}_j$ represents the isospin structure associated with the isovector-vector $\rho$ meson interaction.
Eq. \ref{lagm} includes contributions from baryonic, leptonic, mesonic, and (anti)kaonic degrees of freedom.
In this expression, the fermionic fields $\psi_b$, $\psi^{\nu}_d$, and $\psi_l$ correspond to the baryon octet, $\Delta$ baryons, and leptons, respectively. The quantities $m_b$, $m_d$, $m_K$, and $m_l$ denote the corresponding bare masses of these particles, including the isospin doublet for (anti)kaons.

The AMMs are introduced by the interaction 
of electromagnetic field with the $\sigma_{\mu\nu}=\frac i2 [\gamma_\mu,\gamma_\nu]$
and strength $\kappa_b$. 
The nonlinear contribution arising from scalar self-interactions in the matter sector is commonly incorporated through a potential of the form \cite{1977NuPhA.292..413B,1987ZPhyA.327..295G,1987ZPhyA.326...57G,1982PhLB..114..392G,1985ApJ...293..470G}
\begin{equation}
U(\sigma) = \frac{1}{3}\, g_1\, m_N\, (g_{\sigma N}\sigma)^3 + \frac{1}{4}\, g_2\, (g_{\sigma N}\sigma)^4,
\end{equation}
where $g_1$ and $g_2$ are dimensionless parameters that determine the form of the scalar potential, and the subscript $N$ denotes coupling to nucleons. 
In the expression (\ref{lagm})
\begin{equation}
\omega_{\mu\nu}~=~\partial_\nu\omega_\mu~-~\partial_\mu\omega_\nu,
\end{equation}
\begin{equation}
{\mbox{\boldmath $\rho$}}_{\mu\nu}~=~\partial_\nu{\mbox{\boldmath $\rho$}}_\mu~-
~\partial_\mu{\mbox{\boldmath $\rho$}}_\nu,
\end{equation}
\begin{equation}
F^{\mu\nu} = \partial_\mu A_\nu - \partial_\nu A_\mu.
\end{equation}

In {ZM model \cite{1990PhRvC..42.1416Z}} the Lagrangian density is 
\begin{eqnarray}
{\cal L}_m &=& \sum_b \left[\left(1+\frac{g_{\sigma b}}{m_b}\right)
\bar\psi_{b}\left(i\gamma_\mu 
D^{\mu} -\sigma_{\mu\nu}\kappa_b F^{\mu\nu}
+ g_{\sigma b} \sigma - g_{\omega b} \gamma_\mu \omega^\mu
- g_{\rho b}
\gamma_\mu{\mbox{\boldmath $\tau$}}_b \cdot
{\mbox{\boldmath $\rho$}}^\mu \right)\psi_b
\right. \nonumber\\
&& \left. - \bar\psi_{b} m_b \psi_b \right]
 + \frac{1}{2} \partial_\mu \sigma\partial^\mu \sigma
- \frac12 m_\sigma^2 \sigma^2
 \nonumber\\
&& -\frac{1}{4} \omega_{\mu\nu}\omega^{\mu\nu}
+\frac{1}{2}m_\omega^2 \omega_\mu \omega^\mu
- \frac{1}{4}{\mbox {\boldmath $\rho$}}_{\mu\nu} \cdot
{\mbox {\boldmath $\rho$}}^{\mu\nu}
+ \frac{1}{2}m_\rho^2 {\mbox {\boldmath $\rho$}}_\mu \cdot
{\mbox {\boldmath $\rho$}}^\mu \nonumber \\
&& + \sum_{l=e,\mu} \bar\psi_{l}\left(i\gamma_\mu 
D^{\mu} - m_l \right)\psi_l, 
\label{lagmzm}
\end{eqnarray}
with the covariant derivative, $D^\mu=\partial^\mu + ieQA^\mu$.
The difference between
Walecka model and ZM model is that in non-linear Walecka
model $g_1\neq0$, $g_2\neq0$ and
\begin{equation}
m_b^*~=~m_b\left(1-\frac{g_{\sigma b} \sigma}{m_b}\right)
\end{equation}
but in ZM model $g_1=0$, $g_2=0$ and
\begin{equation}
m_b^*~=~m_b\left(1+\frac{g_{\sigma b} \sigma}{m_b}\right)^{-1},
\end{equation}
where $\sigma$ is given by its ground state expectation value
\begin{equation}
<\sigma>=\sigma~=~\frac1{m_\sigma^2}\left(\sum_b g_{\sigma b} ~ n_b^{(S)}
~-~\frac{\partial U}{\partial \sigma}\right)
\end{equation}
and $n_b^{(S)}= \langle\bar\psi_b\psi_b\rangle$
is the scalar density.

The electromagnetic field Lagrangian density is given by
\begin{equation}
{\cal L}_f  =  -\frac1{16\pi} F_{\mu \nu}F^{\mu \nu}, \label{lagf}
\end{equation}

At vanishing temperature, and assuming a constant magnetic field, the scalar, number, and kinetic energy densities of neutral baryons are obtained as \cite{2000ApJ...537..351B,2001ApJ...546.1126S,2008JPhG...35l5201R}
\begin{equation}
n_b^{(S)}~=~\frac{m_b^*}{(2\pi)^2} \sum_s \left[p_{F,s}^{(b)} E_F^{(b)}
-\bar{m_b}^2\ln\left(\frac{p_{F,s}^{(b)}+E_F^{(b)}}{\bar{m_b}}\right)\right],
\label{scdennt}
\end{equation}

\begin{equation}
n_b~=~\frac1{2\pi^2} \sum_s \left[\frac13 p_{F,s}^{(b)^3}
+ \frac12 s \kappa_b B \left[\bar{m_b} p_{F,s}^{(b)}
+  E_F^{(b)^2}
\left(\arcsin \frac {\bar{m}}{E_F^{(b)}} - \frac{\pi}{2} \right)
\right]\right]
\end{equation}
and

\begin{eqnarray}
\varepsilon_b = \frac1{(4\pi)^2} \sum_s
\left[p_{F,s}^{(b)} E_F^{(b)^3} + \frac83 s \kappa_b B E_F^{(b)^3}
\left(\arcsin \frac {\bar{m_b}}{E_F^{(b)}} - \frac{\pi}{2} \right)
\right. \nonumber \\
\left. +\left(\frac43 s \kappa_b B - \bar{m_b} \right)
\left[\bar{m_b} p_{F,s}^{(b)} E_F^{(b)}
+ \bar{m_b}^3\ln\left(\frac{p_{F,s}^{(b)}+E_F^{(b)}}{\bar{m_b}}\right)\right]\right].
\label{endennt}
\end{eqnarray}
For charged baryons the scalar, number and energy densities are
given by \cite{2000ApJ...537..351B,2001ApJ...546.1126S,2008JPhG...35l5201R}

\begin{equation}
n_{b'}^{(S)}~=~\frac{e|Q|B}{2\pi^2} \sum_{n=0}^{n_{max}}\sum_s 
m_{b'}^*\frac{\sqrt{m_{b'}^{*^2} + 2ne|Q|B} + s\kappa_{b'}B}
{\sqrt{m_{b'}^{*^2} + 2ne|Q|B}}
\ln\left(\frac{p_{F,n,s}^{(b')}+E_F^{(b')}}
{\sqrt{m_{b'}^{*^2} + 2ne|Q|B} + s\kappa_{b'}B}\right),
\end{equation}

\begin{equation}
n_{b'}~=~\frac{e|Q|B}{2\pi^2} \sum_{n=0}^{n_{max}} \sum_s  p_{F,n,s}^{(b')}
\end{equation}
and

\begin{eqnarray}
\varepsilon_{b'} =  
\frac{e|Q|B}{2\pi^2}  \sum_{n=0}^{n_{max}}\sum_s
\left[p_{F,n,s}^{(b')}E_F^{(b')}
~+~\left(\sqrt{m_{b'}^{*^2} + 2ne|Q|B} + s\kappa_{b'}B\right)^2 \right.
\nonumber \\ \left.  \times  
\ln \left(\frac{p_{F,n,s}^{(b')}+E_F{b')}}
{\sqrt{m_{b'}^{*^2} + 2ne|Q|B} + s\kappa_{b'}B}\right) \right].
\label{endench}
\end{eqnarray}
For leptons the the number and energy densities are \cite{1997PhRvL..78.2898C,1998PhRvD..58l1301B,2000ApJ...537..351B,2001ApJ...546.1126S,2007MPLA...22..623C,2008JPhG...35l5201R} 

\begin{equation}
n_l~=~\frac{e|Q|B}{2\pi^2} \sum_{n=0}^{n_{max}} (2-\delta_{n,0})  p_{F,n}^{(l)}
\end{equation}
and

\begin{equation}
\varepsilon_l =
\frac{e|Q|B}{(2\pi)^2}  \sum_{n=0}^{n_{max}}
(2-\delta_{n,0}) \left[p_{F,n}^{(l)}E_F^{(l)}~+~(m_l^2 + 2ne|Q|B) 
\ln \left(\frac{p_{F,n}^{(l)}+E_F^{(l)}}{\sqrt{m_l^2 + 2ne|Q|B}}\right)\right].
\label{endenl}
\end{equation}
In the Eqs. (\ref{scdennt}-\ref{endenl}) we have used the short
hand notations $\bar{m} = m^* + s \kappa B$, $p_{F,s} = \sqrt{E_F^2 - \bar{m}^2}$
and $p_{F,n,s} = \sqrt{E_F^2 -\left(\sqrt{m^{*^2} + 2ne|Q|B} - s\kappa B\right)^2}$,
$E_F$ being the fermi energy of the respective particle.
The total energy density of the matter sector, derived from the Lagrangian density ${\cal L}_m$ as specified in Eqs.~(\ref{lagm}) and (\ref{lagmzm}), can be written as
\begin{eqnarray}
\varepsilon_m = \sum_b \varepsilon_b + \sum_{b'} \varepsilon_{b'} 
+ \sum_{l}\varepsilon_l 
+\frac12 m_\sigma^2 \sigma^2 + U(\sigma)
+ \frac12 m_\omega^2 \omega^{0^2}+\frac12 m_\rho^2 \rho_3^{0^2},
\label{enden}
\end{eqnarray}
Here, the indices $b$ and $b'$ correspond to neutral and charged baryons, respectively, while $l$ represents the leptonic contributions.
The associated matter pressure follows from standard thermodynamic considerations and can be expressed as
$P_m=\sum_b \mu_b n_b + \sum_l \mu_l n_l - \varepsilon_m$,
where the chemical potential of each species is defined as $\mu_i = \partial \varepsilon_m / \partial n_i$.

\subsection{Numerical results - for nucleon ($n,p,e^-$) matter}

The most acceptable and the most simplistic model of highly 
dense matter inside the NS core assumes the matter 
is composed of neutrons, protons and electrons in beta-equilibrium.
Nucleon matter in strong magnetic field has been studied in
many literature \cite{2000ApJ...537..351B,2001ApJ...546.1126S,
2008JPhG...35l5201R,1997PhRvL..78.2898C,1998PhRvD..58l1301B,
2007MPLA...22..623C}. % + hyperon references

For nucleon matter protons and electrons are subjected to Landau
quantization. Since baryons are dominating factor for dense matter,
the contribution from Landau quantization comes when the protons
occupy only the zeroth Landau level. The charge neutrality condition
$n_p=n_e$ gives

\begin{equation}
\sum_{n=0}^{n_{max}^{(p)}} p_{F,n}^{(p)} = \sum_{n=0}^{n_{max}^{(e)}} p_{F,n}^{(e)}.
\end{equation}
This shows that for any field the number of occupied Landau 
levels are same for both proton and electron for charge neutral 
$n$-$p$-$e$ system. 
For magnetic field strengths satisfying $B \gtrsim B_c^{(e)}$ while remaining well below $B_c^{(p)}$, many Landau levels remain occupied, and the overall behaviour of the system is not significantly different from the non-magnetized case. In contrast, for fields considerably exceeding $B_c^{(e)}$, the quantization effect becomes pronounced for electrons, leading to the occupation of only a few Landau levels. Under the requirement of charge neutrality, this reduction in electron phase space also influences the proton distribution in a similar manner \cite{1997PhRvL..78.2898C}.
This happens for magnetic
field strength $B\ge5\times 10^{18}$ G \cite{2000ApJ...537..351B}.
Below this field the difference from the field free case is
not significant \cite{2000ApJ...537..351B, 1997PhRvL..78.2898C}.

\begin{figure*}[t!]
\begin{center}
%\resizebox{0.75\textwidth,90\angle}{!}{
\includegraphics[width=7.5 cm, keepaspectratio]{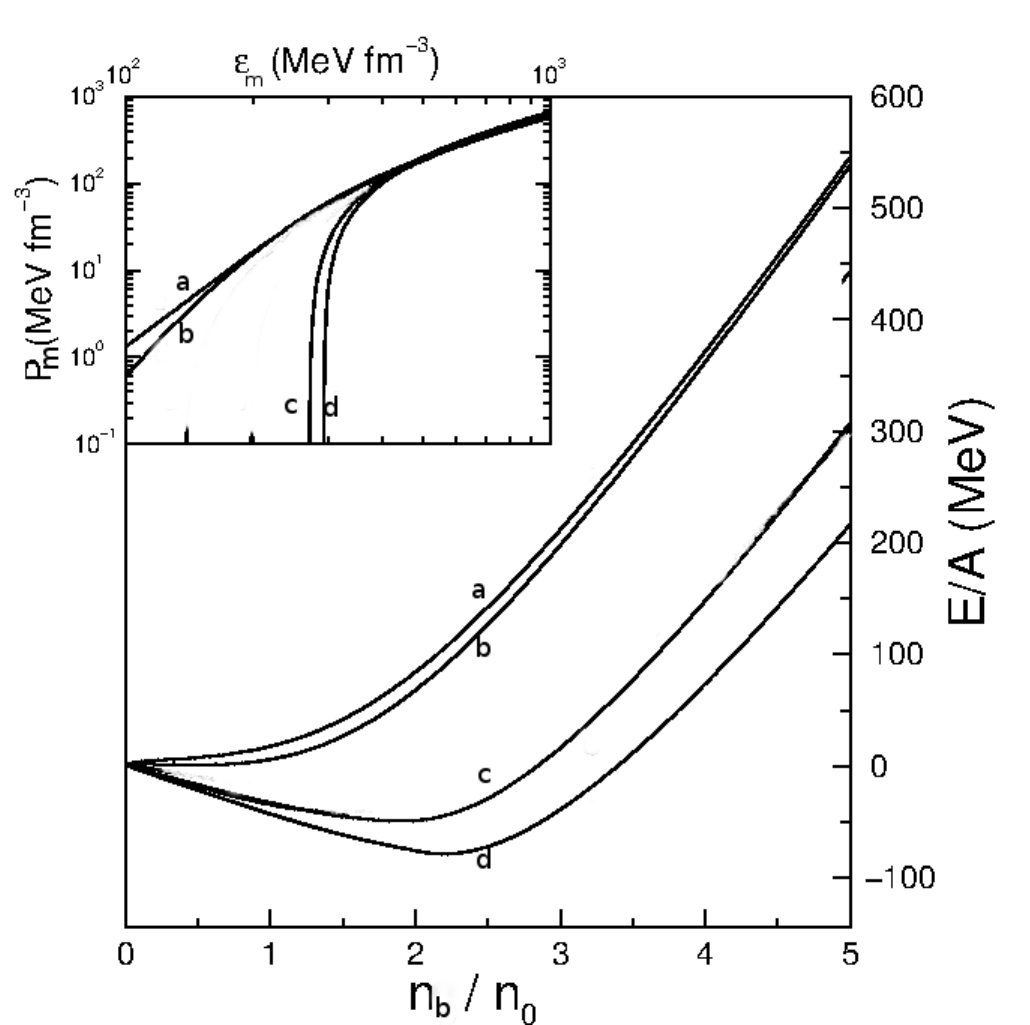}
%}
%\vspace*{5cm}
\caption{Model HS81: variation of binding energy per nucleon 
of nucleonic matter with model HS81 with the normalized baryonic 
density for different magnetic fields. The curve $a$ is without 
magnetic field while the curves $b$, $c$ and $d$ are for $B=10^4B_c^{(e)},
~B_c^{(p)}~{\rm and}~10B_c^{(p)}$ respectively. The pressure 
as a function of energy density is shown in the inset. The 
result has been taken from \cite{1997PhRvL..78.2898C}.} 
\label{deba1}
\end{center}
\end{figure*}

At this stage, it is useful to briefly clarify the nuclear model employed in the calculations. The HS81 parameter set \cite{1981NuPhA.368..503H} refers to a RMF description developed to reproduce the bulk properties of symmetric nuclear matter at saturation density. 
In this model, the interaction between nucleons is mediated through scalar and vector meson fields, and the parameters are chosen to yield realistic values of binding energy, effective mass and incompressibility. The HS81 parametrization is often used as a representative stiff EoS and provides a useful baseline for examining the influence of external magnetic fields on nucleonic matter.
The notable point is that while in absence of magnetic field 
$n$-$p$-$e$ system is not a bound state, sufficiently strong 
magnetic field makes it bound \cite{1997PhRvL..78.2898C} as 
shown in Fig. \ref{deba1} for model HS81.
The $n$-$p$-$e$ system shows no binding in the absence of
magnetic field (curve $a$). When the field is slightly quantizing
($B=10^4B_c^{(e)}$) the matter still remain unbound which is 
shown by curve $b$. However the matter becomes bound when 
the electrons are strongly quantized which is evident from 
the curves $c$ and $d$ for $B=B_c^{(p)}$ and $B=10B_c^{(p)}$ 
respectively.

\begin{figure*}[h!]
\begin{center}
%\resizebox{0.75\textwidth,90\angle}{!}{
\includegraphics[width=6.6 cm, keepaspectratio]{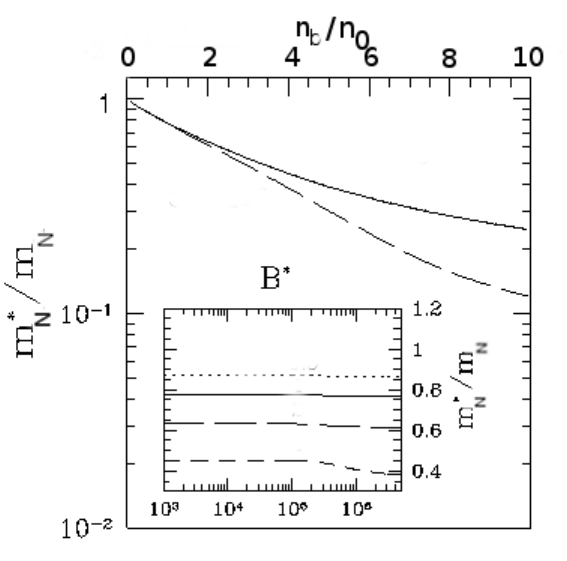}
\hspace{.3cm}
\includegraphics[width=6.6 cm, keepaspectratio]{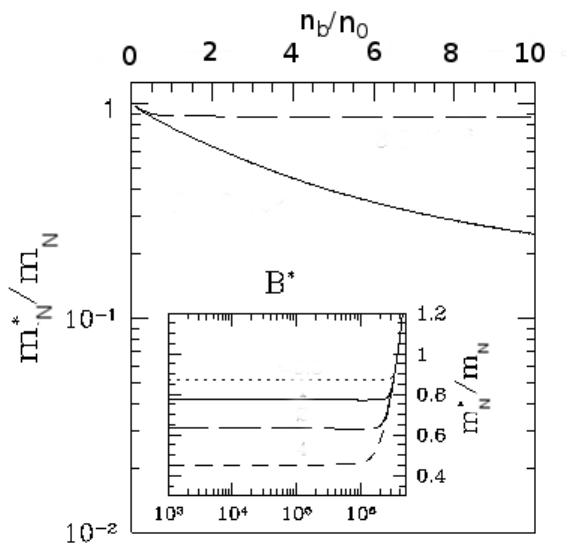}
%}
\caption{Model GM3: variation of $m^*/m$ for nucleonic 
matter with the normalized baryonic density for
different field strength. Dotted curve is for $B=0$, solid 
line for $B=10^5B_c^{(e)}$, dashed line for $B=3.3\times10^6B_c^{(e)}$.
Inset: the variation of $m^*/m$ with field strength for different
baryon density. The dotted curve is for $n_B/n_0 = 0.5$, solid
line for $n_B/n_0 = 1$, long dashed curve for $n_B/n_0 = 2$,
short dashed for $n_B/n_0 = 4$. Left panel: $m^*/m$ without 
AMM, right panel: with AMM. The result has been taken from
\cite{2000ApJ...537..351B}.}
\label{bpl20001}
\end{center}
\end{figure*}

Another widely used parametrization within the RMF framework is the GM3 model \cite{1991PhRvL..67.2414G}. This set of parameters is constructed to reproduce empirical nuclear matter properties and has been extensively applied in NS studies due to its relatively moderate stiffness. Compared to HS81, the GM3 model generally predicts a different density dependence of the effective mass and pressure, making it particularly suitable for examining how magnetic field effects interplay with the underlying nuclear interaction. The use of both HS81 and GM3 models therefore allows us to assess the model dependence of the results presented in this work.
The comparison between these two models also helps in identifying generic features that arise due to magnetic field effects, independent of the specific choice of nuclear interaction.
The variation of the effective mass ratio 
of nucleons as a function of the normalized baryon density for different magnetic field strengths within the GM3 parametrization is illustrated in Fig. \ref{bpl20001}. 
The left panel corresponds to the case without inclusion of AMM effects, while the right panel includes these contributions. It is observed that, in the absence of AMM, the magnetic field leads to a moderate modification in the effective mass primarily through Landau quantization. However, when AMM effects are incorporated, the deviation becomes more pronounced, especially at higher densities and stronger fields. The inset further highlights the sensitivity of the effective mass to the magnetic field at fixed densities, indicating that the influence of the field becomes increasingly significant with increasing baryon density. This behaviour reflects the interplay between scalar mean-field interactions and magnetic effects in determining the in-medium properties of baryons.

\begin{figure*}[h!]
\begin{center}
%\resizebox{0.75\textwidth,90\angle}{!}{
\includegraphics[width=7.5 cm, keepaspectratio]{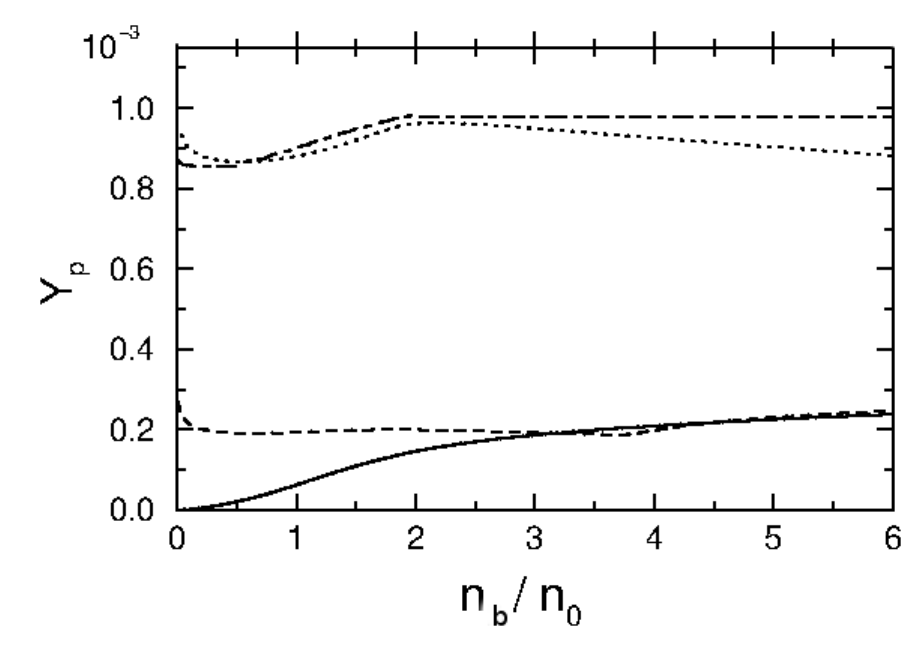}
%}
\caption{Model HS81: variation of proton fraction with the 
normalized baryonic density for different magnetic fields.
The solid curve is for $B=0$, the dashed curve for $B=10^4B_c^{(e)}$,
the dotted curve for $B=B_c^{(p)}$ and the dot-dashed curve
for $B=10B_c^{(p)}$ respectively. The result has been taken 
from \cite{1997PhRvL..78.2898C}.}
\label{deba3}
\end{center}
\end{figure*}

\begin{figure*}[h!]
\begin{center}
%\resizebox{0.75\textwidth,90\angle}{!}{
\includegraphics[width=6.6 cm, keepaspectratio]{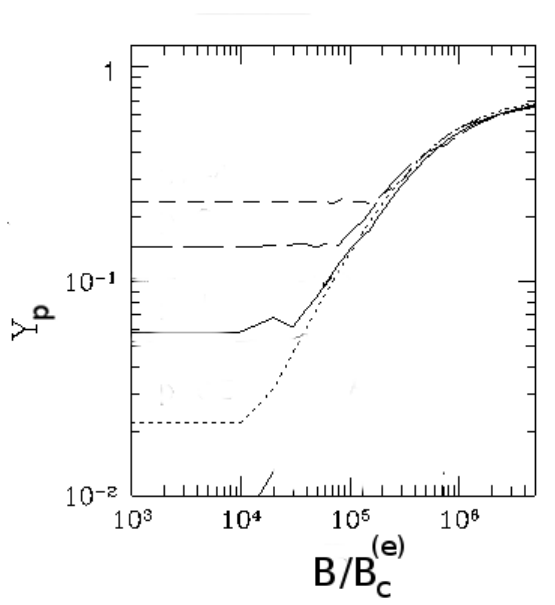}
\hspace{.3cm}
\includegraphics[width=6.7 cm,height=7.2cm]{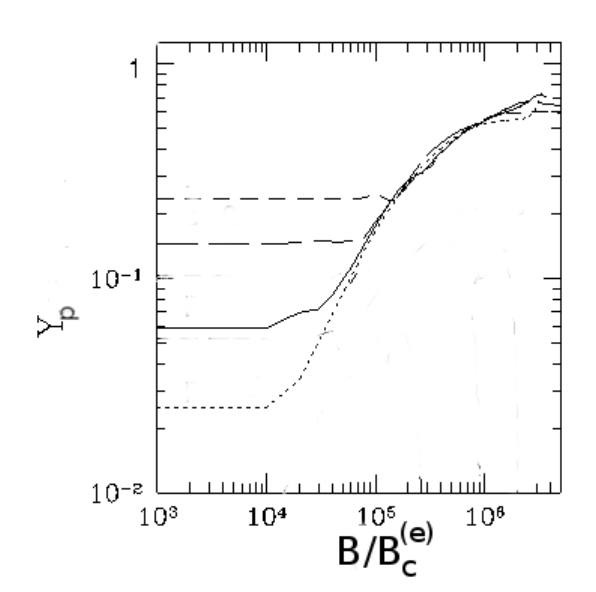}
%}
\caption{Model GM3: variation of proton fraction in nucleonic
matter with the field strength for different baryonic density
Left panel: Proton fraction without AMM, right 
panel: with AMM. The dotted curve is for $n_B/n_0=0.5$, the
solid line for $n_B/n_0=1$, the long dashed curve for $n_B/n_0=2$
and the short-dashed curve for $n_B/n_0=4$. The result has 
been taken from \cite{2000ApJ...537..351B}.}
\label{bpl20003}
\end{center}
\end{figure*}

A strong magnetic field substantially alters the composition 
of matter. For field strength $10^4B_c^{(e)}$ proton fraction 
enhances significantly at the low density regime and becomes 
almost independent of density. The feature has been shown in 
Figs. \ref{deba3} and \ref{bpl20003} for the model HS81 and
GM3 respectively. The proton fraction saturates with the further 
increase of field strength when field strength is near and 
above $B_c^{(p)}$. However, if nucleon AMM
is considered the fraction of spin polarized neutron increases 
with the increase of magnetic field strength \cite{2000ApJ...537..351B,
2003ChPhL..20.1238M}. However, this becomes significant only at
very high field strength.

One important feature of the NS observation is its
thermal evolution with time or more specifically the cooling
of NS. The cooling of NS at the early stage 
is mostly dominated by neutrino emission mainly from the core
of the star, which is the result of various internal processes. 
For non-superfluid matter modified URCA and Bremstrahlung processes
occur through out the core while for massive stars direct
URCA and Bremstrahlung processes occur only in the inner core
of the star where proton fraction is above some threshold.
These two direct processes are responsible for fast cooling.
In presence of magnetic field the threshold density for occurring
direct URCA processes is smeared out and become lower compared to
cases without magnetic field.

The most powerful neutrino emission process inside the neutron
star is direct URCA process
\begin{equation}
n \rightarrow p\,+\,e\,+\,\bar{\nu}_e,\,\,\, p\,+\,e\, \rightarrow n +\,\nu_e.
\nonumber
\end{equation}
if the process is operative. Inside NS the matter 
is degenerate. Hence the particles may take part in any reaction 
only if their momenta are in the vicinity of their respective 
Fermi momenta. This leads to the constrain for direct URCA such as
\begin{equation}
p_{Fp}\,+\,p_{Fe}\,\ge \,p_{Fn}.
\nonumber
\end{equation}
This demands proton fraction inside NS above some
threshold density which is possible only for higher matter density
at the inner core of massive stars. Consequently, the direct URCA
process does not occur for low massive star inside which the
matter density and hence the proton fraction do not reach the
threshold value for direct URCA to occur. Hence, neutrino cooling
process is via slow modified URCA and Bremsstrahlung processes 
except highly massive stars where fast neutrino emission processes
occur inside inner core. However, in presence of magnetic field
the situation is different. First as already mentioned before, 
the presence of magnetic field enhances the proton fraction
increasing the possibility to open the direct URCA and direct 
Bremsstruhlung processes. Second, in the allowed region magnetic
field enhances the neutrino emissivity 
compared to that in field free case \cite{1998PhRvD..58l1301B}.
This behaviour can be qualitatively understood from the modification of the phase space and kinematic conditions induced by the magnetic field. 
The Landau quantization of charged particles restricts their transverse motion and alters the density of states, effectively relaxing the strict momentum conservation constraints that govern weak interaction processes in the absence of a magnetic field. 
As a result, the direct URCA process can operate over a wider range of densities. 
The combined effect of modified kinematics and the enhanced proton fraction leads to an increase in neutrino emission rates by about $1-2$ orders of magnitude compared to the non-magnetized case.
For detailed reviews of NS cooling processes, see Refs. \cite{2004ARA&A..42..169Y,2006NuPhA.777..497P}.

\begin{figure}
\begin{center}
%\resizebox{0.75\textwidth,90\angle}{!}{
\includegraphics[width=6.5 cm,keepaspectratio]{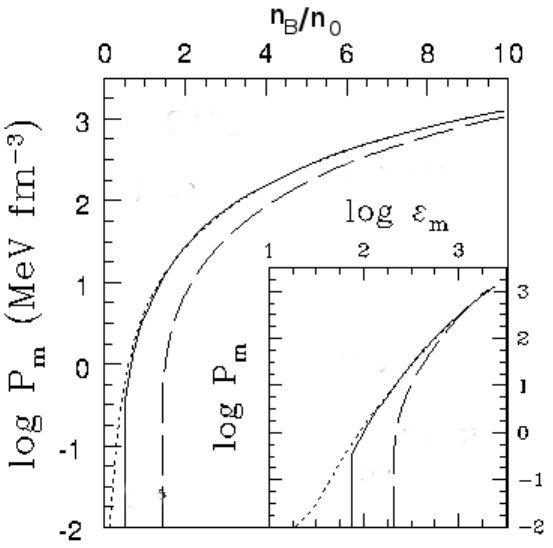}
\hspace{.3cm}
\includegraphics[width=6.5 cm,keepaspectratio]{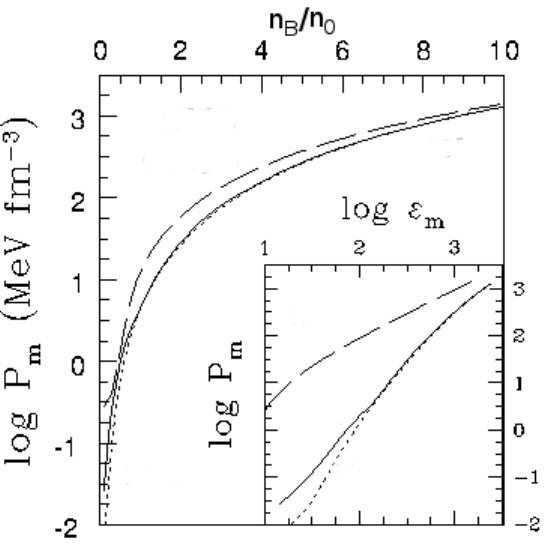}
%}
\caption{Model GM3: variation of pressure for nucleonic matter 
with the normalized baryonic density for different field strength.
Inset: the variation of pressure with energy density baryon 
density.  Dotted curve is for $B=0$, solid line for $B=10^5B_c^{(e)}$, 
dashed line for $B=3.3\times10^6B_c^{(e)}$. Left panel: EoS 
wihout AMM, right panel: with AMM. The result has been taken 
from \cite{2000ApJ...537..351B}.}
\label{bpl20002}
\end{center}
\end{figure}

Landau quantization makes the EoS softer 
\cite{2000ApJ...537..351B,2008JPhG...35l5201R,1997PhRvL..78.2898C,2007MPLA...22..623C}. 
The behaviour is independent of model. Left panel of Fig. 
\ref{bpl20002} shows the softening of EoS with increase of 
field strength for the model GM3. The variation of pressure 
with energy density for different field strength is shown in 
insets of Fig. \ref{deba1} and left panel of Fig. 
\ref{bpl20002} for models HS81 and GM3 respectively.
On the other hand AMM effect is applicable to protons and neutrons.
Contrary to the effect of Landau quantization, AMM effect stiffens 
the EoS \cite{2000ApJ...537..351B,2008JPhG...35l5201R}.
It should be noted that the effect of AMM is more significant
than those of Landau quantization at $B\ge5\times 10^{18}$ G 
\cite{2000ApJ...537..351B}. Consequently, at such a strong
magnetic field the complete description of dense matter must
include AMM effect with Landau quantization. At large magnetic
field the stiffening due to AMM effect overwhelms the softening
effect of Landau quantization as shown in the right panel of 
Fig. \ref{bpl20002} compared to the left panel of the figure.
However, the pressure shown here is only the matter pressure,
the contribution from the field pressure has not been taken into
account.

\begin{figure*}[t!]
\begin{center}
%\resizebox{0.75\textwidth,90\angle}{!}{
\includegraphics[width=6.47 cm,keepaspectratio]{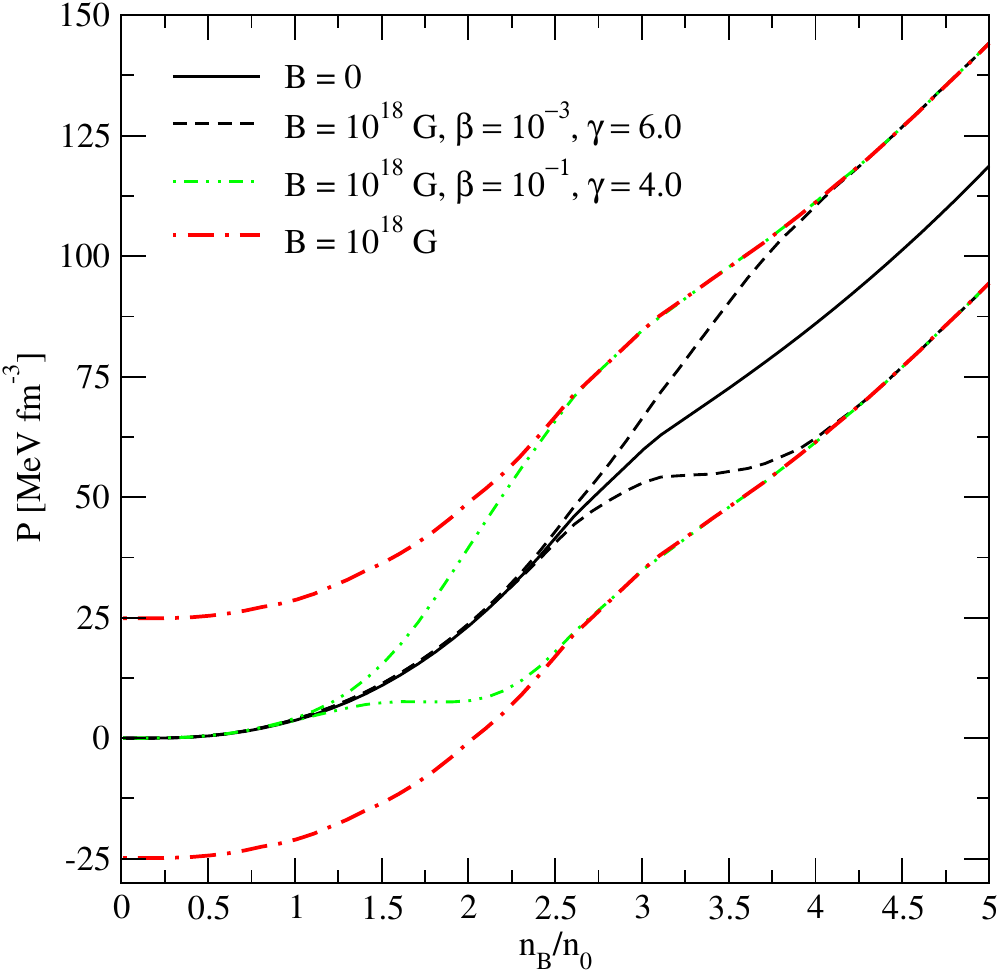}
\hspace{.3cm}
\includegraphics[width=6.5 cm,height=6.25cm]{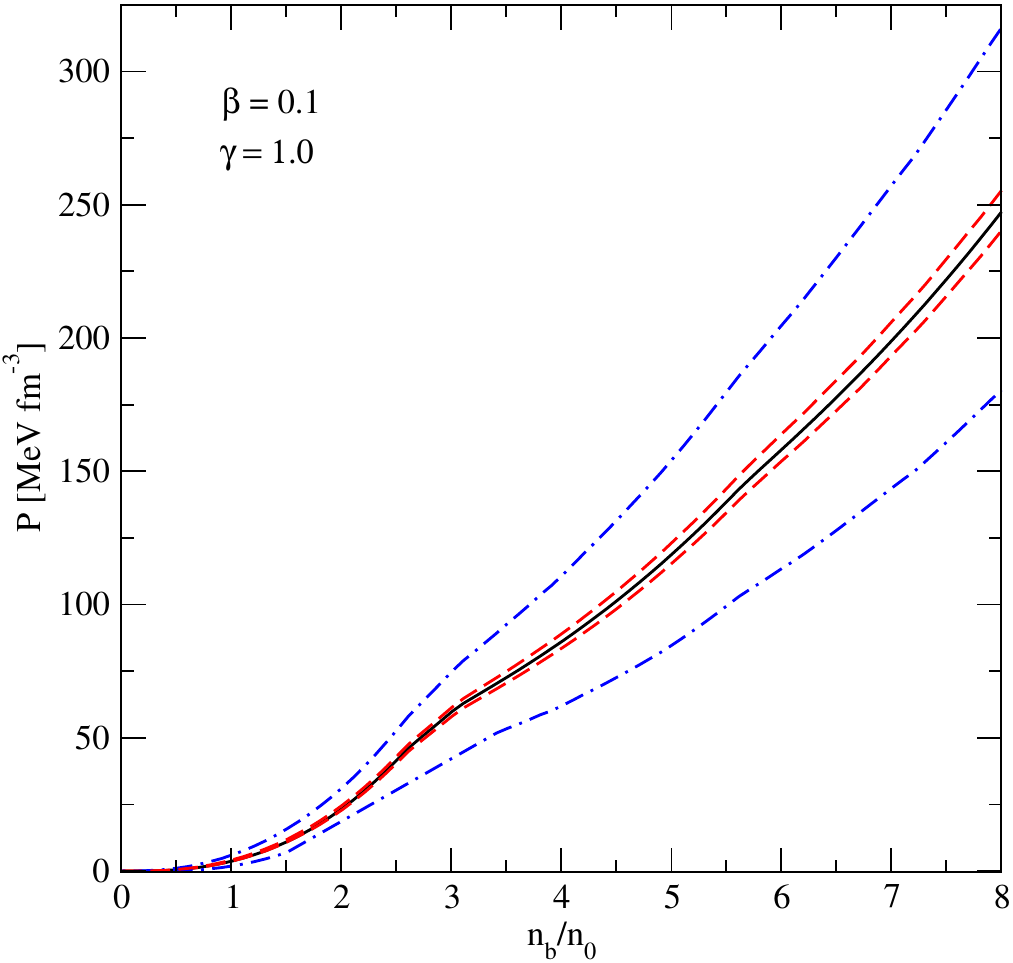}
%}
\caption{Dependence of the total pressure on the baryon number density normalized to saturation density. 
\textit{Left panel:} Results are shown for two representative central magnetic field strengths, $B_c=0$ and $B_c=10^{18}$ G, using different magnetic field profiles. The profiles correspond to parameter sets $\beta=10^{-3}$, $\gamma=6$ (dashed lines), $\beta=10^{-1}$, $\gamma=4$ (double-dashed lines), and the limiting case $\beta \rightarrow \infty$, representing a constant field $B_c$ (dash-dotted lines). For each set of curves, the upper branch denotes the perpendicular pressure component $P_\perp$, while the lower branch corresponds to the parallel component $P_\parallel$. 
\textit{Right panel:} The same quantities are displayed for a fixed profile characterized by $\beta=10^{-1}$ and $\gamma=1$, considering two different central field strengths, $B_c=10^{18}$ G (dashed lines) and $B_c=3\times10^{18}$ G (dash-dotted lines). The field-free case ($B_c=0$) is indicated by the solid curve. The results are adapted from Ref.~\cite{2013NuPhA.898...43S}.}
\label{asymmp-n}
\end{center}
\end{figure*}

%\subsection{Hyperonic matter}
\subsection{Magnetic Field profile}

\begin{figure}[t!]
    \centering
    \includegraphics[width=7.0cm,keepaspectratio]{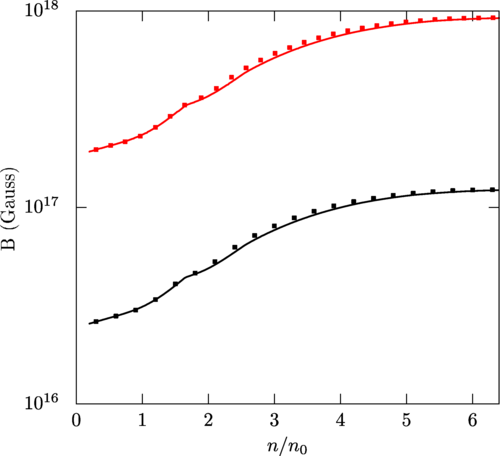}
    \caption{Magnetic field as a function of the normalized density $n/n_0$ for the profile given by Eq. (\ref{eq.45}). The upper curves correspond to $\mu=1.5 \times 10^{32}\rm{A}-\rm{m}^2$ and the lower curves to $\mu=2 \times 10^{31}\rm{A}-\rm{m}^2$. Solid and dotted lines indicate the DD-ME2 and DD-MEX parametrizations, respectively. Figure adapted from \cite{2023PhRvC.107c5807K}.}
    \label{fig:7}
\end{figure}

It should be noted here that most of the studies shows the
effect of magnetic field on the matter pressure except a few \cite{2013NuPhA.898...43S,2010PhRvC..82f5802F,2011PhRvD..83d3009P}. These literature studied the structure of star considering the influence of 
strong magnetic field to the total pressure and energy density with other contribution coming directly from 
presence of magnetic field, not only affecting the single particle 
properties. In these old literature, it was stated that in presence of magnetic field the energy density 
is enhanced by the factor $B^2/8\pi$ as shown in the Eq. (\ref{ener0}) 
while the total pressure becomes anisotropic as evident from 
Eqs. (\ref{perp}) and (\ref{par}). The field pressure contributes 
positively to matter pressure in direction of field enhancing 
the pressure in that direction while it reduces the total pressure 
in the direction normal to field contributing in opposite direction. 
This makes the matter unstable above some critical value of 
field strength near about $10^{19}$ G \cite{2013NuPhA.898...43S}. 
This limits the magnetic field strength inside NS. 
The anisotropicity of pressure in presence of magnetic field 
has been discussed for hypernuclear and quark matter in many 
studies \cite{2013NuPhA.898...43S,2010PhRvC..82f5802F,2011PhRvD..83d3009P}. In the light of this theory,
Fig. \ref{asymmp-n} shows how magnetic field affects the 
EoS where the variation of field strength within the star has 
been considered. 
%{\color{red}\subsection{Magnetic Field profile}
The field profile inside the star is assumed 
as
%In the above discussion of numerical results for hyperonic matter, 
%the field profile inside the star {\color{red} has been assumed as}
\begin{equation}
B\left(\frac{n_b}{n_0}\right)=B_s+B_c\left\{1-\exp\left[-\beta
\left(\frac{n_b}{n_0}\right)^\gamma\right]\right\}
\end{equation}
where $B_s$ and $B_c$ are surface and central field strength
respectively and $\beta$, $\gamma$ are two parameters as proposed 
by \cite{1997PhRvL..79.2176B}. 
In more recent works \cite{2022PhRvC.106c5801M,2023ApJ...943...52R,2025PhRvD.111f3030C}, magnetic field distributions inside NSs are often modeled using profiles derived from solutions of the coupled Einstein–Maxwell equations rather than purely phenomenological density-dependent parametrizations. A commonly used approach is based on fitting numerical magnetostatic equilibrium solutions obtained for different equations of state. One such profile corresponds to a polynomial fit of the monopolar component of the magnetic field and can be written as \cite{2019PhRvC..99e5811C}
\begin{equation}
    B(x)=B_m (1-1.6x^2-x^4+4.2x^6-2.4x^8),
\end{equation}
where $x=r/r_{\rm mean}$ and $B_m$ is the central field strength. While this form captures the general increase of the magnetic field toward the stellar interior, it is better suited for simplified geometries. A more realistic prescription incorporates a poloidal field structure by expressing the magnetic field as a function of the baryon chemical potential. In this case, the field is parametrized as \cite{2017PhLB..773..487D}
\begin{equation}\label{eq.45}
    B(\mu_B)=(a+b\mu_B+c\mu_B^2)\left(\frac{\mu}{B^{(e)}_c}\right),
\end{equation}
where $\mu_B$ is the baryon chemical potential (in units of MeV), $\mu$ is the stellar magnetic dipole moment (in units of Am$^2$) and $B^e_c=4.414\times 10^{13}$ G, is the critical electron field. Such profiles are constructed from quadratic fits to Einstein–Maxwell solutions and ensure a smooth variation of the magnetic field across the stellar interior, even in the presence of phase transitions. For typical dipole moments consistent with \textit{magnetar} observations, these models yield surface fields of order $10^{16}$ G and central fields reaching $10^{17}-10^{18}$ G.
Importantly, these parametrizations describe the spatial distribution of the magnetic field while keeping the thermodynamic pressure isotropic; any anisotropy in the total stress tensor arises solely from the electromagnetic field contribution and not from the intrinsic matter EoS.
The behaviour of such parametrizations is illustrated in Fig. \ref{fig:7}, where the magnetic field is seen to increase progressively with baryon density, reaching its maximum toward the stellar core, with the detailed profile depending on the chosen model parameters.

It should be noted that anisotropy comes from field stresses, not from the EoS.
The confusion often arises when spatial components of the energy–momentum tensor are
interpreted directly as pressures. The EoS provides a scalar pressure p, but the
full energy–momentum tensor also contains contributions from the electromagnetic field.
The electromagnetic field itself carries energy and momentum. Its energy–momentum
tensor is intrinsically anisotropic. For a magnetic field pointing along a particular direction
(say the z-direction), the field produces magnetic pressure perpendicular to the field lines
and magnetic tension along the field lines. This leads to direction-dependent stresses in the
spatial components of the total energy–momentum tensor.
Thus, the anisotropy observed in energy momentum tensor is due to electromagnetic field stresses and exists
even in vacuum. It is not a property of the matter EoS. The thermodynamic
pressure of matter remains isotropic; only the total stress tensor becomes anisotropic due
to the presence of the magnetic field.

\subsection{Hyperonic matter and $\Delta-$admixed hypernuclear matter}

For massive stars if the matter density at inner core of neutron
star is high enough, there is the possibility to appear hyperons
if the fermi momenta of constituent particles at that high
density becomes larger than the mass of the hyperons depending
upon some other conditions. The effect of strong magnetic field
on the matter composed of nucleons and hyperons has also been
studied in several literature in the frame work of RMF model and ZM model with different parametrization 
\cite{1999ApJ...525..950Y,2002PhLB..531..167B,2009PhRvC..79b5803Y,
2010PhRvC..82b5804R,2013NuPhA.898...43S}.
With the strange hyperons, there is also the possibility of appearance of heavier non-strange baryons $\Delta$ isobars. Recent covariant density functional calculations incorporating the full baryon octet together with the $\Delta$(1232) resonance quartet in strong magnetic fields provide new insight into the interplay between $\Delta-$resonances and hyperons in dense matter \cite{2020Parti...3..660T, 2021EPJA...57..216D}.
Within density-dependent RMF models with DD-ME2 parametrization, it was shown that Landau quantization of charged baryons and leptons modifies the population thresholds of both hyperons and $\Delta$ states, leading to non-trivial oscillatory behaviour in particle fractions and effective masses.
A key result is that the inclusion of $\Delta-$resonances shifts the onset of hyperons to higher densities, thereby stiffening the high-density part of the EoS. In strong magnetic fields (up to central values $B_c \sim 10^{18}$ G), this interplay becomes more intricate due to the Landau quantization of charged $\Delta$ states and hyperons.
The occupation of successive Landau levels induces de Haas–van Alphen–type oscillations in the partial densities and in the Dirac effective mass.

\begin{figure}[t!]
    \centering
    \includegraphics[width=9.0cm,keepaspectratio]{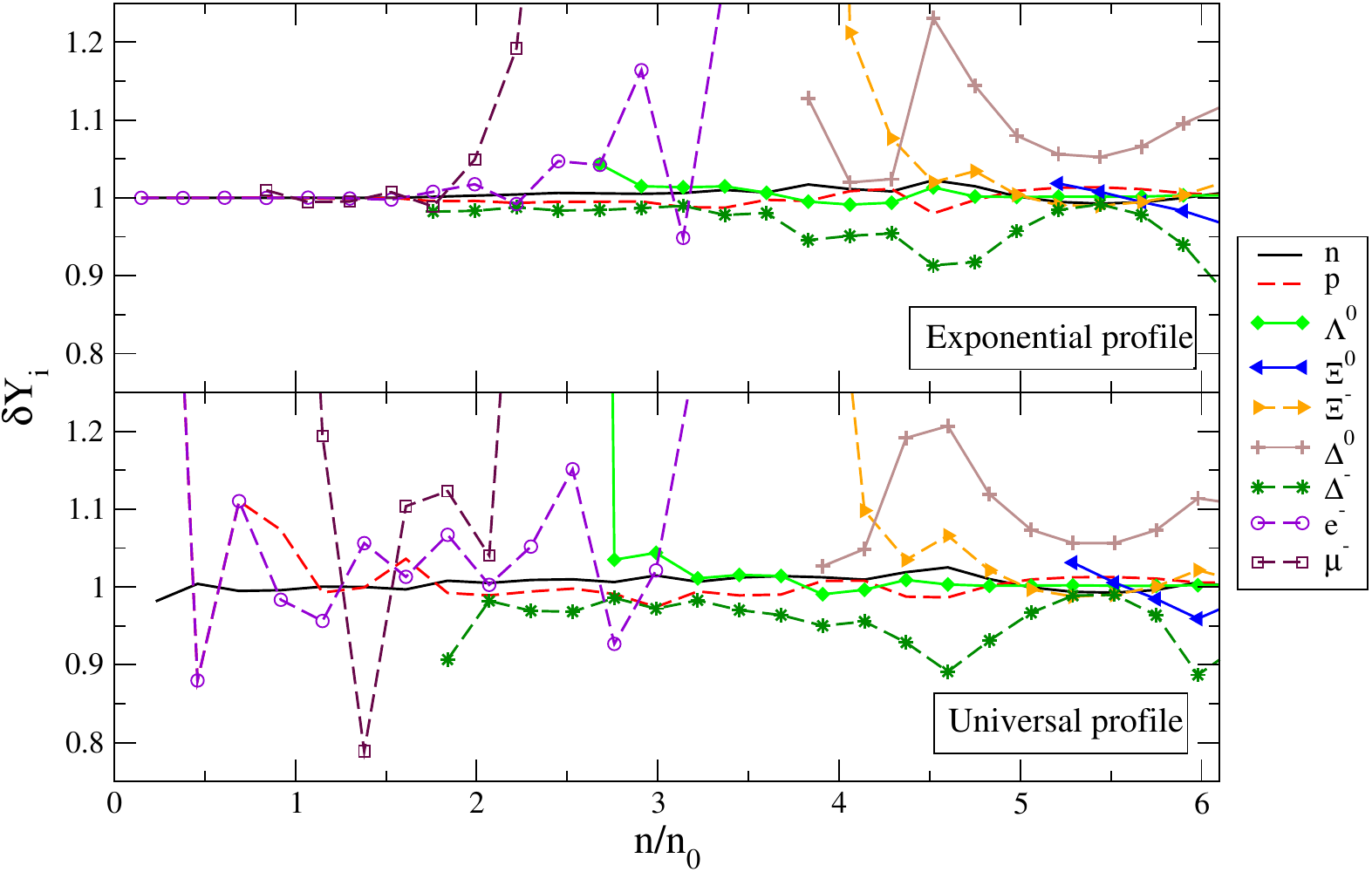}
    \caption{Variation of $\delta Y_i=n_i(B)/n_i(0)$ with normalized baryon density $n/n_0$ for baryon octet and $\Delta$-baryons. The right panel indicates the particle markers. Upper and lower panels represent exponential and universal magnetic-field profiles, respectively.
    Figure adapted from \cite{2020Parti...3..660T}.}
    \label{fig:MDPI-EOSB-fraction}
\end{figure}

Interestingly, although magnetic fields generally stiffen the EoS, the quantitative increase in the maximum mass remains modest. For a central field, $B_c \sim 2.9 \times 10^{18}$ G, the increase in the maximum mass for $\Delta$-admixed hyperonic matter is of the order of $\sim 0.01\% - 0.03\%$, depending on the adopted magnetic-field profile.
This indicates that while magnetic fields significantly affect microscopic composition (up to $\sim 4\%$ changes in hyperon fractions), their macroscopic impact on stellar global parameters is comparatively small for realistic field configurations.
Another important aspect concerns charge neutrality. The Landau quantization enhances the electron fraction in strong fields, which suppresses negatively charged $\Delta^-$ populations. This modifies the delicate balance between $\Delta^-$ and hyperons, thereby altering the strangeness content and isospin composition of matter (please refer to Fig.-\ref{fig:MDPI-EOSB-fraction}). 
The strange sector is found to be more sensitive to magnetic fields than the non-strange sector, implying that magnetic fields may influence hyperon-driven cooling and transport properties more strongly than previously appreciated.

\subsection{Boson Condensates}

Bosonic condensation in dense matter under strong magnetic fields exhibits qualitatively different behaviour depending on the spin, isospin, and coupling structure of the mesonic mode. In compact-star conditions, negatively charged antikaons ($K^-$) and charged pions ($\pi^-$) are the most frequently discussed candidates.
While both may condense once their in-medium excitation energy equals the appropriate chemical potential, the underlying mechanisms and magnetic-field sensitivities differ substantially.

\subsubsection{Antikaon Condensation}

\begin{figure}
    \centering
    \includegraphics[width=9.2 cm, keepaspectratio]{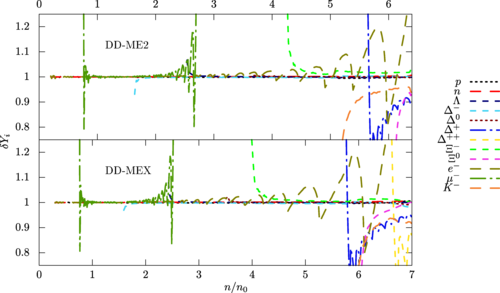}
\hspace{1.0cm}
\includegraphics[width=9.2 cm, keepaspectratio]{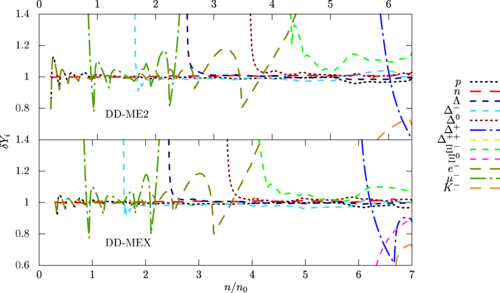}
%}
\caption{Variation of $\delta Y_i=n_i(B)/n_i (0)$ as a function of the total baryon density normalized to saturation density, for a stellar dipole moment $\mu=2 \times 10^{31}\rm{A}-\rm{m}^2$ (upper panel) and $\mu=1.5 \times 10^{32}\rm{A}-\rm{m}^2$ (lower panel), respectively. Figure adapted from \cite{2023PhRvC.107c5807K}.}
    \label{fig:antikaonPRC-fraction}
\end{figure}

Within the density-dependent relativistic mean-field (DD-RMF) framework developed in Ref. \cite{2023PhRvC.107c5807K} (see also references therein), antikaon condensation proceeds predominantly in the $s-$wave channel. In a uniform magnetic field $B$ directed along the $z-$axis, the charged kaon dispersion relation reads
\begin{equation}
    \omega_K^2=p^2_z+m^{*2}_K+|q_{K^-}|B
\end{equation}
with in-medium effective mass,
$m^{*2}_K=m^2_K-g_{\sigma K}\sigma,$ and effective chemical potential
$\mu_{K^-}=\sqrt{m^{*2}_K}+|q_{K^-}|B-g_{\omega K}\omega_0 - \frac{1}{2}g_{\rho K}\rho_{03}$.
It is also pointed out in Ref. \cite{2023PhRvC.107c5807K} that the delayed appearance of antikaons due to the inclusion of magnetic field reduces the softening of the EoS and leads to a slight increase in the maximum NS mass (at the level of $\sim 0.05\%$ for realistic surface fields).
The appearance density of $K^-$ is shifted toward higher baryon densities (typically around $\sim 5.5~n_0$ depending on the parametrization), leading to a mild stiffening of the high-density EoS since the softening associated with the Bose-condensed phase occurs at larger densities.
This behaviour is illustrated in Fig.-\ref{fig:antikaonPRC-fraction}, where the particle population and the threshold density of $K^-$ condensation are shown to shift to higher densities under such conditions of strong magnetic fields.
Thus, in compact-star conditions, magnetic fields primarily shift the threshold density of $s$-wave antikaon condensation without inducing qualitative instabilities.

\subsubsection{Pion Condensation} 

In contrast, pion condensation exhibits a richer structure. As discussed extensively in Ref. \cite{2025PhRvD.111c6022V} and references therein, charged pions in external magnetic fields behave as relativistic charged scalar fields with self-interaction, described schematically by a $\lambda \phi^4$ model.
In a magnetic field and rotating frame, the Klein–Gordon equation acquires additional Coriolis and Landau terms.
Unlike the kaon case, rotation can lower the effective excitation energy and potentially trigger vacuum instabilities at sufficiently large angular velocities, $\Omega$.
Moreover, Ref. \cite{2025PhRvD.111c6022V} shows that rapid rotation can produce large-winding-number ``supervortices'' and induce instabilities even when the static vacuum remains stable.

\section{Quark Matter}\label{sec:quark}
At high baryon densities, the interior of NSs may experience a transition from hadronic matter to a deconfined quark phase. In this context, it has been suggested that quark matter consisting of roughly equal proportions of {\it u}, {\it d}, and {\it s} quarks could represent a stable state even at zero temperature and vanishing pressure \cite{1984PhRvD..30..272W}.
This concept, often known as the strange quark matter hypothesis, suggests that compact stars could exist in which matter is entirely composed of deconfined quarks. Such objects may form either during the collapse of massive stars or through later evolutionary processes of NSs. In certain cases, quark matter may be confined to the central region, surrounded by hadronic matter, giving rise to hybrid star configurations. In contrast, stellar objects consisting entirely of strange quark matter are typically classified as strange stars.

The structural properties of strange stars differ markedly from those of conventional NSs. In particular, their mass–radius relation allows for compact configurations with relatively small radii even at low masses. The precise characteristics depend sensitively on the model employed to describe quark matter. Among the various theoretical approaches, the MIT bag model \cite{1974PhRvD...9.3471C} has been widely used due to its simplicity, treating quarks as quasi-free particles confined within a finite region. Early investigations of strange quark matter were largely based on this framework \cite{1984PhRvD..30.2379F,1986ApJ...310..261A,1986A&A...160..121H}. However, this model does not fully incorporate essential features of quantum chromodynamics, such as chiral symmetry breaking at low densities. To address these limitations, alternative descriptions involving density-dependent quark masses have been developed, which effectively incorporate aspects of confinement and chiral symmetry restoration. Several parameterizations for the density dependence of quark masses have been proposed in the literature \cite{1991PhRvD..43..627C,1998PhLB..438..123D,2000PhRvC..62a5204W,2001EL.....56..361Z,2002PhRvC..65c5202Z,2003PhRvC..67a5202Z,2010MNRAS.402.2715L,2013PhRvD..88b5008S}.

The influence of strong magnetic fields on strange quark matter has also been examined in a number of studies \cite{1996PhRvD..54.1306C,2008PhRvC..77a5807G,2010PhRvC..82f5802F,2011PhRvD..83d3009P,2016PhRvC..94e5805F}. Compared to electrons, quarks carry smaller electric charges and possess larger effective masses, which implies that significantly higher magnetic field strengths are required for Landau quantization to dominate their dynamics. Estimates indicate that the critical field is of the order of $10^{16}$ G for {\it u} and {\it d} quarks, and can reach $\sim10^{19}$ G for {\it s} quarks \cite{2008PhRvC..77a5807G}. Consequently, the strange quark sector is only weakly affected by magnetic fields typically expected in NSs. Substantial modifications to the properties of strange quark matter arise only at very high field strengths approaching $\sim10^{19}$ G \cite{1996PhRvD..54.1306C}. When AMM effects are included, noticeable changes in the pressure may occur at somewhat lower fields, of the order of $\sim10^{18}$ G \cite{2008PhRvC..77a5807G}. Beyond the bag model, various alternative descriptions of magnetized quark matter have also been explored \cite{2013LNP...871...87G,2013LNP...871..399F}.

In scenarios where hadronic and quark matter coexist, forming a mixed phase inside compact stars, the presence of a magnetic field does not significantly alter the phase boundary under equilibrium conditions \cite{1997PhRvL..79.2176B}. However, strong magnetic fields can influence the dynamics of phase conversion. For instance, magnetic-field–induced effects may facilitate the nucleation of quark matter, enabling the transition from hadronic matter under suitable conditions \cite{1996PhRvD..54.1306C,2001IJMPD..10...89G}. Once initiated, such a conversion can propagate through the star, potentially transforming it partially or entirely into a strange star via a shock-driven mechanism \cite{2006PhRvC..74f5804B}. While magnetic fields can assist in triggering the conversion, they may also reduce the propagation speed of the conversion front.

In addition to unpaired quark matter, color-superconducting phases have attracted considerable attention in recent years. Strong magnetic fields influence these phases not only through Landau quantization but also by modifying pairing patterns and stability conditions in color–flavor–locked (CFL) matter. These changes can affect the overall energetics of quark matter and, consequently, the conditions under which such phases may appear in compact stars or form self-bound configurations. A detailed study of magnetized CFL matter and its implications for quark star structure can be found in Ref.~\cite{2010PhRvC..82f5802F}.

\section{Dark Matter in Magnetized Compact Stars}\label{sec:darkmatter}

\subsection{Two-Fluid Formalism for Dark Matter–Admixed Stars}
Compact stars have recently emerged as promising astrophysical laboratories to probe the properties of dark matter (DM), particularly in scenarios where DM interacts with hadronic matter only gravitationally \cite{2025Univ...11...74G}. 
In such models, NSs may accumulate a non-negligible DM component either through gravitational capture or via inheritance from DM-rich progenitor environments. A consistent theoretical framework for studying these systems is provided by the two-fluid formalism, where the hadronic matter (HM) and DM sectors satisfy independent hydrostatic equilibrium equations but contribute jointly to the spacetime curvature through Einstein’s equations \cite{2025MNRAS.543...83K}.

\begin{figure}[t!]
    \centering
    \includegraphics[width=0.95\linewidth]{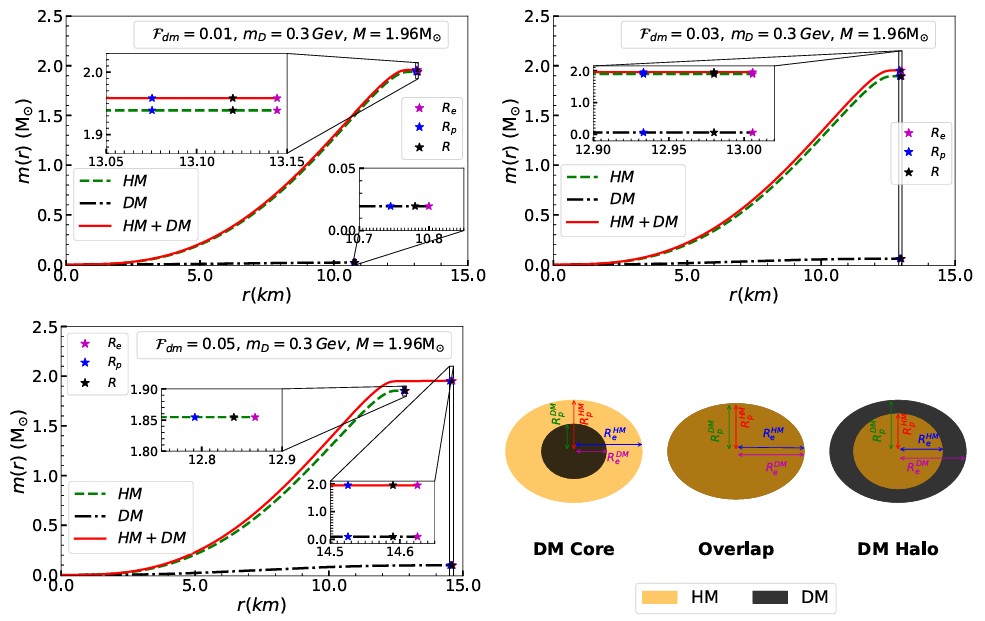}
    \caption{Radial mass distribution of hadronic and dark matter components in magnetized dark matter–admixed NSs, illustrating core, mixed, and halo configurations. Figure adapted from Ref.-\citep{2025MNRAS.543...83K}.}
    \label{fig:dm_core_halo}
\end{figure}

Within this approach, the TOV equations generalize to coupled structure equations for the two fluids. 
Depending on the relative stiffness of the hadronic and DM EoSs, several configurations may arise: (i) a DM core embedded within hadronic matter, (ii) a mixed configuration where both fluids share the same surface, or (iii) a DM halo extending beyond the visible stellar surface. 
The radial mass distribution of the hadronic and dark matter components for these different configurations is illustrated in Fig.-\ref{fig:dm_core_halo}.
The latter configuration is particularly intriguing, as it allows the presence of an extended gravitationally bound DM envelope that does not participate in electromagnetic emission but nonetheless modifies the external spacetime geometry \cite{2018PhRvD..97l3007E}.

\subsection{Magnetic Deformation and Non-Symmetric Dark Matter Halos}
The interplay between ultra-strong magnetic fields and DM-admixed NSs introduces qualitatively new effects. In magnetars, where central magnetic fields may reach $10^{17}-10^{18}$ G, the magnetic pressure anisotropy deforms the stellar structure from spherical symmetry \cite{2010PhRvC..82f5802F}. 
When treated perturbatively within general relativity, the magnetic field modifies both the mass distribution and the spacetime curvature.
Although DM does not couple directly to the magnetic field, it responds to the magnetically deformed spacetime through gravitational interaction \cite{2009APh....32..278S,2011PhLB..695...19C}.
Following a Hartle-type perturbative treatment, the total energy–momentum tensor is written as \cite{1967ApJ...150.1005H},
\begin{equation}
 T_{\mu \nu}=T_{\mu \nu}^0+\delta T_{\mu \nu},   
\end{equation}
where the perturbation includes magnetic pressure contributions. The magnetic pressure perturbation can be expanded in spherical harmonics as
\begin{equation}
    \delta P=p_0(r) +p_2(r)P_2(\cos \theta),
\end{equation}
with $P_2(\cos \theta)=(3\cos^2 - 1)/2$.
The monopole perturbation equations are expressed as \cite{2014PhRvC..89d5805M}
\begin{equation}
    \frac{dm_0}{dr}=4\pi r^2 p_0, \quad
    \frac{dh_0}{dr}=4 \pi r e^{2\lambda}p_0+\frac{d\nu}{dr}\frac{e^{2\lambda}m_0}{r}+ \frac{e^{2\lambda}m_0}{r^2}.
\end{equation}

A particularly novel outcome of this mechanism is the possibility of a non-symmetric DM halo. In configurations where the DM radius exceeds that of the hadronic component, the magnetically induced quadrupolar deformation propagates into the DM sector, leading to direction-dependent distortions of the halo. The equatorial and polar radii of the DM component may differ by tens to hundreds of meters for magnetar-strength fields. This effect becomes more pronounced for stiff DM EoS and larger DM fractions \cite{2025MNRAS.543...83K}.

Importantly, such non-symmetric halos can generate observational signatures distinct from those of ordinary \textit{magnetars}. Since electromagnetic observations probe only the visible hadronic surface, the presence of an extended DM halo can lead to discrepancies between the inferred stellar radius and the total gravitational mass. Moreover, the quadrupolar deformation of the halo modifies the external gravitational field, potentially affecting pulsar timing observables such as the period $P$ and its derivative $\dot{P}$, as well as the amplitude of continuous gravitational-wave emission.

\begin{figure}[h!]
    \centering
    \includegraphics[width=0.75\linewidth]{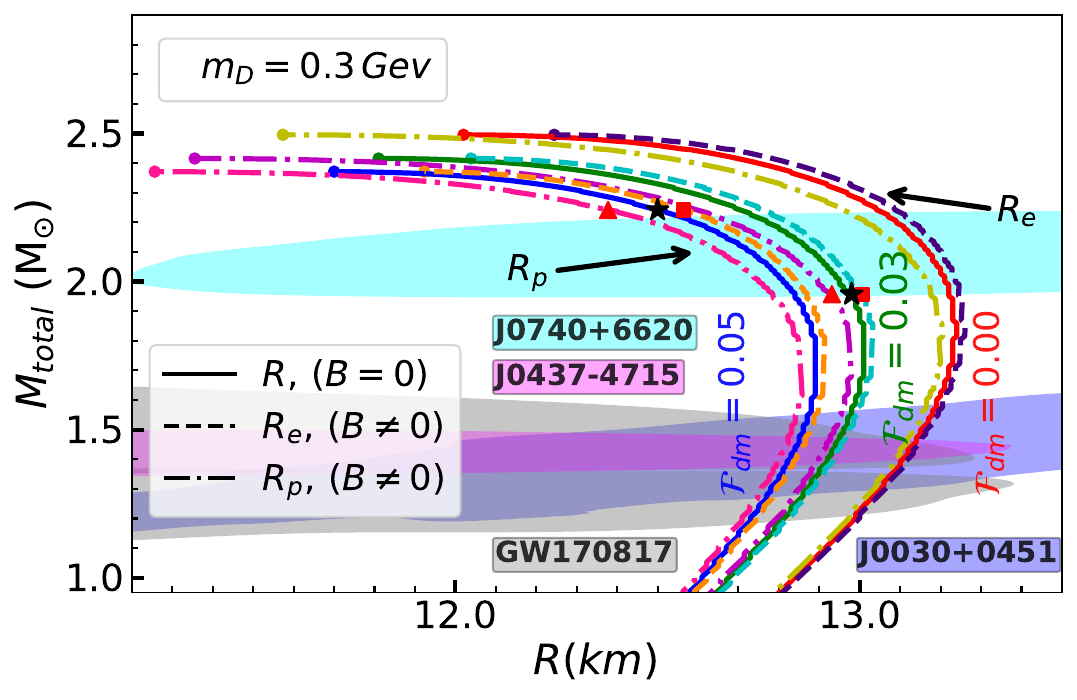}
    \caption{Mass–radius relation for magnetized dark matter–admixed NSs for different dark matter fractions. The polar and equatorial radii illustrate magnetic deformation, while observational constraints from NICER and GW170817 are shown for comparison. Figure adapted from Ref.-\citep{2025MNRAS.543...83K}.}
    \label{fig:mr_dm}
\end{figure}

\subsection{Impact on Structure and Observational Signatures}
The presence of a DM component in magnetized NSs modifies both their equilibrium structure and potential observational signatures. In two-fluid frameworks where DM interacts with hadronic matter only gravitationally \cite{2009APh....32..278S,2011PhLB..695...19C}, the effective EoS is generally softened at high densities (for fermionic DM), leading to slightly more compact stars and a modest reduction in the maximum mass \cite{2011PhRvD..84j7301L,2018PhRvD..97l3007E,2020JPhG...47i5202Q}. 
The corresponding mass–radius relation for magnetized dark matter–admixed NSs, together with observational constraints from NICER and gravitational-wave measurements, is presented in Fig.-\ref{fig:mr_dm}.
For realistic DM fractions $(\leq 5\%)$, the resulting mass–radius relations remain consistent with current NICER and gravitational-wave constraints.
However, sufficiently stiff DM EoSs or larger DM fractions can produce halo configurations in which the DM radius exceeds that of the visible HM surface \cite{2019JCAP...07..012N}. 
In magnetars, ultra-strong magnetic fields induce quadrupolar spacetime deformations that increase the gravitational mass perturbatively and distort the stellar geometry \cite{2023PhRvD.108h3003P}. 
Although DM does not couple directly to the magnetic field, it responds to the magnetically deformed curvature, potentially forming non-symmetric halos with direction-dependent extensions. Such asymmetric DM envelopes alter the external gravitational multipole structure without affecting electromagnetic emission from the visible surface, thereby introducing possible discrepancies between the inferred radius and total gravitational mass. These effects may influence pulsar timing parameters and continuous gravitational-wave strain, making magnetars promising laboratories for probing gravitational signatures of dark matter in compact stars.

\section{Neutron star structure}\label{sec:NS-structure}

The presence of a magnetic field introduces directional dependence in the energy–momentum tensor, which in turn challenges the assumption of spherical symmetry underlying the TOV equations. As a consequence, strongly magnetized NSs are expected to deviate from spherical configurations and exhibit structural deformations. The impact of magnetic fields on NS structure has been extensively investigated in the literature \cite{1995A&A...301..757B,2001ApJ...554..322C,2001MNRAS.327..639I,2005MNRAS.359.1117T,2014PhRvC..89d5805M,2015MNRAS.447.3785C}.
While a number of studies have focused on the modification of the EoS through microscopic mechanisms such as Landau quantization, magnetization, and pressure anisotropy, it is equally important to examine how these effects influence the global stellar configuration. In this context, the use of spherically symmetric TOV equations becomes inadequate for sufficiently strong magnetic fields, motivating the development of axisymmetric treatments based on the coupled Einstein–Maxwell equations.

Assuming alignment between the magnetic and rotational axes, and considering predominantly poloidal field configurations, it has been shown that magnetic stresses tend to elongate the star along the symmetry axis, resulting in an oblate deformation \cite{1995A&A...301..757B,2001ApJ...554..322C,2014PhRvC..89d5805M,2015MNRAS.447.3785C}. Similar qualitative behaviour has also been reported in models that include toroidal field components.

A comprehensive and self-consistent analysis was presented in Ref.~\cite{2015MNRAS.447.3785C}, where magnetized NSs were studied within a fully relativistic, axisymmetric framework. This work quantified the relative contributions of microscopic magnetic effects and large-scale electromagnetic stresses in determining stellar structure. It was found that the dominant influence on the global configuration arises from the Lorentz force rather than from the gravitational contribution of the electromagnetic stress-energy tensor \cite{1995A&A...301..757B}.

Early studies suggested that the inclusion of strong magnetic fields could increase the maximum mass of NSs by approximately $13$–$29\%$, depending on the underlying EoS \cite{1995A&A...301..757B,2001ApJ...554..322C}. However, more recent self-consistent calculations, which incorporate magnetic-field–dependent equations of state, magnetization effects, and full axisymmetric solutions of the Einstein–Maxwell system, indicate that these enhancements are considerably smaller. Even for extreme interior field strengths approaching $10^{18}$–$10^{19}$ G, the impact on global properties such as mass and radius remains relatively modest.
The combined effects of rotation and magnetic fields have also been explored. In rapidly rotating configurations with mixed poloidal–toroidal magnetic fields, the resulting stellar shapes remain predominantly oblate \cite{2005MNRAS.359.1117T}. However, for observed magnetars, which typically have rotational periods of the order of one second, rotational effects are expected to be subdominant compared to magnetic-field–induced deformations \cite{2001ApJ...554..322C}.

\section{Summary and Outlook}\label{sec:summary}

In this review, we have discussed the role of strong magnetic fields in modifying the microscopic and macroscopic properties of dense matter relevant to compact stars. The presence of intense magnetic fields introduces several nontrivial effects, most notably Landau quantization and anomalous magnetic moment interactions, which significantly influence the energy spectra, composition, and thermodynamic properties of dense matter. These effects, in turn, alter the equation of state and thereby affect the global structure and observable properties of NSs.

We have examined the behaviour of magnetized hadronic matter within RMF frameworks, including the possible emergence of additional degrees of freedom such as hyperons, $\Delta$ resonances, and meson condensates. The interplay between these constituents under strong magnetic fields remains a complex problem, particularly due to competing effects such as softening from Landau quantization and stiffening from AMM contributions. Furthermore, the inclusion of magnetic field contributions to the energy–momentum tensor introduces anisotropic stresses, which may have important implications for the stability and deformation of compact stars.

We have also highlighted how strong magnetic fields can influence particle populations, neutrino emission processes, and cooling mechanisms in NSs. In particular, the enhancement of proton fraction and modification of phase space due to Landau quantization can facilitate fast neutrino emission processes such as direct URCA, thereby impacting the thermal evolution of NSs. Despite significant progress, a consistent and unified description of dense matter under extreme magnetic fields that simultaneously incorporates all relevant physical effects remains an open challenge.

In addition to these aspects, recent developments in the study of strongly magnetized matter have revealed the importance of quantum anomaly-driven phenomena and related transport effects \cite{2015ARNPS..65..193K,2026PrPNP.14604199A}. In systems where an imbalance between left- and right-handed fermions is present \cite{PhysRevD.91.061301}, magnetic fields can induce nontrivial transport behaviour, leading to the generation of currents along the field direction \cite{PhysRevD.73.045006,2013LNP...871.....K}. Such mechanisms, often associated with chiral magnetic and related effects, have been extensively investigated in the context of quantum chromodynamics, heavy-ion collisions, and compact astrophysical objects. These effects arise from the underlying symmetry structure of the theory and may influence transport properties, magnetic field evolution, and observable signatures in NSs and magnetars. Moreover, strong magnetic fields are now understood to affect not only bulk thermodynamic quantities but also the phase structure of strongly interacting matter, potentially giving rise to new phases and anomalous transport coefficients.

A systematic incorporation of these emerging effects into existing models of dense matter remains an important direction for future work. In particular, connecting microscopic quantum phenomena with macroscopic astrophysical observables in a consistent framework is essential for advancing our understanding of magnetized compact stars. Continued progress in this area, supported by developments in both theoretical modeling and multi-messenger observations, is expected to provide deeper insights into the physics of dense matter under extreme conditions.

%%%%%%%%%%%%%%%%%%%%%%%%%%%%%%%%%%%%%%%%
%%%%%%%%%%%%%%%%%%%%%%%%%%%%%%%%%%%%%%%%%%
\vspace{6pt} 

%%%%%%%%%%%%%%%%%%%%%%%%%%%%%%%%%%%%%%%%%%
%% optional
%\supplementary{The following supporting information can be downloaded at:  \linksupplementary{s1}, Figure S1: title; Table S1: title; Video S1: title.}


% Only used for preprtints:
% \supplementary{The following supporting information can be downloaded at the website of this paper posted on \href{https://www.preprints.org/}{Preprints.org}.}

%\begin{tabularx}{\textwidth}{lll}
%\toprule
%\textbf{Name} & \textbf{Type} & \textbf{Description} \\
%\midrule
%S1 & Python script (.py) & Script of python source code used in XX \\
%S2 & Text (.txt) & Script of modelling code used to make Figure X \\
%S3 & Text (.txt) & Raw data from experiment X \\
%S4 & Video (.mp4) & Video demonstrating the hardware in use \\
%\bottomrule
%\end{tabularx}
%}

%%%%%%%%%%%%%%%%%%%%%%%%%%%%%%%%%%%%%%%%%%
\authorcontributions{Conceptualization, MS and VBT; methodology, VBT.; writing---original draft preparation, MS and VBT; writing---review and editing, VBT; visualization, VBT; supervision, MS}

\funding{MS gratefully acknowledges the financial support provided by the Science and Engineering Research Board (SERB), Department of Science and Technology, Government of India, under Project No. CRG/2022/000069.}

{\conflictsofinterest{The authors declare no conflict of interest.}}

{\dataavailability{No new data were generated for this study. The findings discussed are based on previously published works, and all relevant sources are cited within the manuscript.}}

\acknowledgments{The authors thank the anonymous referees for their constructive comments which have bestowed to enhance the quality of the manuscript notably.}

%%\conflictsofinterest{The authors declare no conflicts of interest.} 

%%%%%%%%%%%%%%%%%%%%%%%%%%%%%%%%%%%%%%%%%%
%% Optional

%% Only for journal Encyclopedia
%\entrylink{The Link to this entry published on the encyclopedia platform.}

%\abbreviations{Abbreviations}{
%The following abbreviations are used in this manuscript:
%\\

%\noindent 
%\begin{tabular}{@{}ll}
%MDPI & Multidisciplinary Digital Publishing Institute\\
%DOAJ & Directory of open access journals\\
%TLA & Three letter acronym\\
%LD & Linear dichroism
%\end{tabular}
%}

%%%%%%%%%%%%%%%%%%%%%%%%%%%%%%%%%%%%%%%%%%
%% Optional

\reftitle{References}

% Please provide the correct journal abbreviation (e.g. according to the “List of Title Word Abbreviations” http://www.issn.org/services/online-services/access-to-the-ltwa/).

\bibliography{review, review1}

% If authors have biography, please use the format below
%\section*{Short Biography of Authors}
%\bio
%{\raisebox{-0.35cm}{\includegraphics[width=3.5cm,height=5.3cm,clip,keepaspectratio]{Definitions/author1.pdf}}}
%{\textbf{Firstname Lastname} Biography of first author}
%
%\bio
%{\raisebox{-0.35cm}{\includegraphics[width=3.5cm,height=5.3cm,clip,keepaspectratio]{Definitions/author2.jpg}}}
%{\textbf{Firstname Lastname} Biography of second author}

% For the MDPI journals use author-date citation, please follow the formatting guidelines on http://www.mdpi.com/authors/references
% To cite two works by the same author: \citeauthor{ref-journal-1a} (\citeyear{ref-journal-1a}, \citeyear{ref-journal-1b}). This produces: Whittaker (1967, 1975)
% To cite two works by the same author with specific pages: \citeauthor{ref-journal-3a} (\citeyear{ref-journal-3a}, p. 328; \citeyear{ref-journal-3b}, p.475). This produces: Wong (1999, p. 328; 2000, p. 475)

%%%%%%%%%%%%%%%%%%%%%%%%%%%%%%%%%%%%%%%%%%
%% for journal Sci
%\reviewreports{\\
%Reviewer 1 comments and authors’ response\\
%Reviewer 2 comments and authors’ response\\
%Reviewer 3 comments and authors’ response
%}
%%%%%%%%%%%%%%%%%%%%%%%%%%%%%%%%%%%%%%%%%%
%\PublishersNote{}
%\isPreprints{}{% This command is only used for ``preprints''.
%\end{adjustwidth}
\end{document}